\documentclass[centertags,twoside,reqno,11pt]{amsart}

\usepackage{amssymb}
\usepackage{algorithmic}                                    	
\usepackage[linesnumbered,ruled]{algorithm2e} 
\usepackage{amssymb}
\usepackage{amsmath}		
\usepackage{amsfonts}
\usepackage{dsfont}			
\usepackage{float}			
\usepackage{mathrsfs}
\usepackage{mathptmx}
\usepackage{graphicx} 
\usepackage{lineno,hyperref}
\usepackage{pgfplots}
\pgfplotsset{compat=newest}
\usepackage{tikz}
\usepackage{marvosym} 
\usepackage{subcaption}  
\usepackage{graphicx}  
\usepackage{tikz}
 \usetikzlibrary{automata,topaths}
 \usepackage{pgfplots}
 \usepackage{layouts}

 \definecolor{magenta(process)}{rgb}{1.0, 0.0, 0.56}
 \definecolor{firebrick}{rgb}{0.7, 0.13, 0.13}
\definecolor{seagreen}{rgb}{0.18, 0.55, 0.34}
\definecolor{mediumpurple}{rgb}{0.58, 0.44, 0.86}
\definecolor{magenta}{rgb}{1.0, 0.0, 1.0}
\definecolor{mediumpurple}{rgb}{0.58, 0.44, 0.86}
\definecolor{mediumslateblue}{rgb}{0.48, 0.41, 0.93}
\definecolor{darkslateblue}{rgb}{0.28, 0.24, 0.55}
\definecolor{chartreuse(traditional)}{rgb}{0.87, 1.0, 0.0}
\definecolor{chartreuse(web)}{rgb}{0.5, 1.0, 0.0}
\definecolor{burntsienna}{rgb}{0.91, 0.45, 0.32}
\definecolor{red-violet}{rgb}{0.78, 0.08, 0.52}
\definecolor{turquoise}{rgb}{0.19, 0.84, 0.78}
\definecolor{cornflowerblue}{rgb}{0.39, 0.58, 0.93}
\definecolor{steelblue}{rgb}{0.27, 0.51, 0.71}
\definecolor{plum(web)}{rgb}{0.8, 0.6, 0.8}
\definecolor{plum(traditional)}{rgb}{0.56, 0.27, 0.52}
\definecolor{darkpastelpurple}{rgb}{0.59, 0.44, 0.84}

\SetCommentSty{mycommfont}
\usepackage{eurosym}
\usepackage{array}
\usetikzlibrary{arrows,automata, babel,positioning,shapes,calc}
\tikzset{my rect/.style={draw,shape=rectangle, minimum width=1cm, minimum height=1cm}}
\tikzstyle{decision} = [diamond, draw, fill=yellow!20, 
    text width=4.5em, text badly centered, node distance=3cm, inner sep=0pt]
\tikzstyle{block} = [rectangle, draw, fill=blue!20, 
    text width=5em, text centered, rounded corners, minimum height=4em]
\tikzstyle{line} = [draw, -latex']
\tikzstyle{cloud} = [draw, ellipse,fill=red!20, node distance=3cm,minimum height=2em]
\modulolinenumbers[5]

\providecommand{\U}[1]{\protect \rule{.1in}{.1in}}

\newcommand{\be}{\begin{enumerate}}
\newcommand{\ee}{\end{enumerate}}
\newcommand{\bi}{\begin{itemize}}
\newcommand{\ei}{\end{itemize}}
\newcommand{\beqn}{\begin{eqnarray}}
\newcommand{\eeqn}{\end{eqnarray}}
\newcommand{\beq}{\begin{equation}}
\newcommand{\eeq}{\end{equation}}

\usetikzlibrary{patterns}

\pgfdeclareplotmark{BS}{%
  \pgfpathmoveto{\pgfqpoint{-1.2\pgfplotmarksize}{-1.2\pgfplotmarksize}}%
 \pgfpathlineto{\pgfqpoint{-1.2\pgfplotmarksize}{1.2\pgfplotmarksize}}%
 \pgfpathmoveto{\pgfqpoint{1.2\pgfplotmarksize}{-1.2\pgfplotmarksize}}%
  \pgfpathlineto{\pgfqpoint{1.2\pgfplotmarksize}{1.2\pgfplotmarksize}}%
  \pgfpathmoveto{\pgfqpoint{-1.2\pgfplotmarksize}{-1.2\pgfplotmarksize}}%
 \pgfpathlineto{\pgfqpoint{1.2\pgfplotmarksize}{-1.2\pgfplotmarksize}}%
 \pgfpathmoveto{\pgfqpoint{1.2\pgfplotmarksize}{1.2\pgfplotmarksize}}%
  \pgfpathlineto{\pgfqpoint{-1.2\pgfplotmarksize}{1.2\pgfplotmarksize}}%
  \pgfpathmoveto{\pgfqpoint{-1.2\pgfplotmarksize}{-1.2\pgfplotmarksize}}%
  \pgfpathlineto{\pgfqpoint{1.2\pgfplotmarksize}{1.2\pgfplotmarksize}}%
  \pgfpathmoveto{\pgfqpoint{-1.2\pgfplotmarksize}{1.2\pgfplotmarksize}}%
  \pgfpathlineto{\pgfqpoint{1.2\pgfplotmarksize}{-1.2\pgfplotmarksize}}%
  \pgfusepathqstroke
}

\title[Cellular systems coverage probability characterization]{Stochastic geometric modelling and simulation of cellular systems for coverage probability characterization}

\author[H. Nassar]{Hamed Nassar}
\address{Hamed Nassar \hfill\break
	Computer Science Department, Suez Canal University, Ismailia 41522, Egypt}
\email{nassar@ci.suez.edu.eg}

\author[G. Taher]{Gehad Taher}
\address{Gehad Taher \hfill\break
Computer Science Department, Suez Canal University, Ismailia 41522, Egypt}
\email{Gehad.Taher@ci.suez.edu.eg}

\author[S. El-Hady]{El-Sayed El-Hady}
\address{El-Sayed El-Hady \hfill\break
	Basic Sciences Department, Suez Canal University, Ismailia 41522, Egypt}
\email{elsayed\_elhady@ci.suez.edu.eg}

\keywords{Stochastic geometry; IoT; Cellular network;  Downlink;  Uplink; Coverage probability}

\begin{document}




\begin{abstract}
	Stochastic geometry (SG) has been successfully used as a modelling tool for cellular networks to characterize the coverage probability in both the downlink (DL) and uplink (UL) systems, under the assumption that the base stations (BS) are deployed as a Poisson point process. In the present article, we extend this use and provide further results for interference limited and Rayleigh fading networks, culminating in a multifaceted contribution. First, we compactly model the two systems at once, allowing parallels to be drawn and contrast to be created. Also, for DL we manage to obtain two closed form expressions for two special cases. Moreover, for UL, notorious for being difficult, we develop a clever approximation that overcomes the difficulty, yielding excellent results. Additionally, we present two efficient Monte Carlo simulation algorithms, designed primarily to validate the models, but can be of great use for SG modelling of communications systems in general. Finally, we prove two theorems at odds with popular belief in cellular communications research. Specifically, we prove that under the SG model, the coverage probability in both DL and UL is independent of BS density. Based on this revelation, a plethora of results in the literature have to be re-examined to rid them of a parameter that has been proven superfluous.
\end{abstract}
\maketitle

\section{Introduction}
\label{INTRO}

Stochastic geometry was initially stimulated by applications to biology,
astronomy and material sciences, then introduced seriously in the late
nineties to the field of wireless communications \cite{Baccelli10}. It is
particularly suited for modelling large scale wireless communication
networks, where a network is treated as a realization (snapshot) of a
spatial point process in the entire Euclidean plane \cite{Haenggi12}. In
particular, it is a natural approach to describe node locations in randomly
formed networks, e.g. ad hoc networks, or irregular and regular networks,
e.g. cellular networks. It provides a natural way of computing macroscopic
properties, by averaging over the potential geometrical patterns for all
network nodes to obtain key performance characteristics, such as
connectivity, stability, and capacity, as functions of a relatively small
number of parameters, e.g. the intensity of the underlying point process and
the operational parameters. By a spatial average, it is meant an average
calculated over many points in the considered domain. These locations in our
context are the network elements at the time when the snapshot is taken.
Example spatial averages are the fraction of nodes which transmit, the
fraction of space which is covered or connected, the fraction of nodes that
transmit their data successfully, and the average geographic progress
achieved by a node forwarding a packet towards some destination \cite{ElSawy17}.

The most common function of stochastic geometry in wireless communications is to characterize
the signal to interference and noise ratio (SINR), which can be used to
calculate many cellular performance metrics, such as outage probability,
coverage probability, spatial opportunity, spatial throughput, network
throughput, medium access probability and spectral efficiency \cite%
{Okegbile21}, in both downlink (DL) and uplink (UL) directions. It should be
noted, however, that in light the huge of influx of wireless emissions in
recent years, the impact of noise now pales in comparison with interference 
\cite{Chun20}. As such, there is a growing modelling trend (see, for example, \cite{Tang20, Liu20, Haroon20, Kouzayha21}) to replace SINR by SIR,
and we will follow this trend in the present article. In all cases, the most
relevant assumption made when stochastic geometry is used in modelling
cellular networks is that base stations (BS) are deployed in the Euclidean
space as a Poisson Point Process (PPP), although many variant processes,
e.g. Thomas cluster processes or Mat\'{e}rn cluster processes, have been
suggested as well \cite{Blaszczyszyn18}.

The stochastic geometric model of cellular networks, portrayed in Figure \ref%
{fig:Voronoi}, has established itself confidently in the past decade as a
replacement to the once popular hexagonal grid model, in which the base
stations were placed at the centers of the hexagonal lattices. It has been
shown \cite{Lee13} that the PPP approach provides much more accurate results
than the hexagonal grid model when both are used to model real world
cellular installation. It has also been shown \cite{Andrews10} that the PPP
model gives lower bounds, whereas the hexagonal model gives upper bounds, of
the coverage probability, which means that the former is safer to rely on.

\begin{figure}[H]

	\centering
\begin{tikzpicture}[xscale=1.4,yscale=1.4]
	\begin{axis}[
		ticks=none,
		scale only axis,
		xmin=0,
		xmax=8.8,
		ymin=0,
		ymax=7.5,
		axis line style={line width=2pt},
		legend pos=outer north east,
		legend cell align=left,
		legend style={draw=none,font=\small,row sep=0.2cm,/tikz/every odd column/.append style={column sep=0.1cm,text width=8em}},
		]
		
		\addplot[only marks, mark=BS,mark size=2.5pt, red] coordinates{(1.9, 0.68)};
		\addplot[only marks,seagreen,mark size=1.5pt,very thick] coordinates {
			(7.5,0.48)
		}; 
		
		\addplot[mark=BS,mark size=2.5pt, red] coordinates{(6.3, 1.15)};
		
		\addplot[mark=BS,mark size=2.5pt, red] coordinates{(7.1, 0.21)};
		\addplot[mark=BS,mark size=2.5pt, red] coordinates{(1, 3.9)};
		\addplot[mark=BS,mark size=2.5pt, red] coordinates{(0.12, 4.66)};
		\addplot[mark=BS,mark size=2.5pt, red] coordinates{(1.25, 6.77)};
		\addplot[mark=BS,mark size=2.5pt, red] coordinates{(4.9, 2.7)};
		\addplot[mark=BS,mark size=2.5pt, red] coordinates{(7.77, 7)};
		\addplot[mark=BS,mark size=2.5pt, red] coordinates{(8.09, 2.55)};
		
		\addplot[no markers, cyan!70!green] coordinates {
			(0,2)
			(2.6,2.5)
			(4.1,0.94)
			(4.25,0)
		}; 
		\addplot[no markers,cyan!70!green] coordinates {
			(2.6,2.5)
			(3.75,5.15)
			(1.85,5.3)
			(0,3.85)
		}; 
		
		\addplot[no markers,cyan!70!green] coordinates {
			(1.85,5.3)
			(0,5.98)
		}; 
		
		\addplot[no markers,cyan!70!green] coordinates {
			(3.75,5.15)
			(4.6,5.65)
			(4.49,7.5)
		}; 
		
		\addplot[no markers,cyan!70!green] coordinates {
			(4.1,0.94)
			(6.47,2.47)
			(6.62,4.73)
			(4.6,5.65)
		}; 
		
		\addplot[no markers,cyan!70!green] coordinates {
			(6.47,2.47)
			(7.8,1.35)
			(5.6,0)
		}; 
		
		\addplot[no markers,cyan!70!green] coordinates {
			(7.8,1.35)
			(9.5,0.8)
		}; 
		
		\addplot[no markers,cyan!70!green] coordinates {
			(6.62,4.73)
			(9.5,4.9)
		}; 
		
		\addplot[only marks,seagreen,mark size=1.5pt,very thick] coordinates {
			(0.26  ,0.35)
			(0.92 , 0.62)
			(1.84    , 1.57)
			(2.76   ,0.59)
		}; 
		
		\addplot[only marks,seagreen,mark size=1.5pt,very thick] coordinates {
			
			(0.32, 2.7)
			(0.94, 2.52)
			(1.62, 2.75)
			(1.3, 3.25)
			(1.25, 4.51)
			(1.93, 4.8)
			(2.28, 4.25)
			(2.35, 4.75)
			
		}; 
		
		\addplot[only marks,seagreen,mark size=1.5pt,very thick] coordinates {
			
			(1,	5.1)
		}; 
		
		\addplot[only marks,seagreen,mark size=1.5pt,very thick] coordinates {
			
			(1.92, 6.2)
			(3.1, 6.95)
			(4.3, 6.55)
		}; 
		
		\addplot[only marks,seagreen,mark size=1.5pt,very thick] coordinates {
			
			(3.3, 1.98)
			(4.1, 3.38)
			(4.32, 5.15)
			(5.25, 4.92)
			(6.05, 4.45)
			(5.65, 3.6)
		}; 
		
		\addplot[only marks,seagreen,mark size=1.5pt,very thick] coordinates {
			
			(5.6 ,  1.35)
			(6.4 ,  0.89)
			(6.6 ,  2.1)
			
		}; 
		
		\addplot[only marks,seagreen,mark size=1.5pt,very thick] coordinates {
			
			(5.68, 7.1)
			(6.3, 6.25)
			(7.55, 5.72)
			(7.75, 6.55)
			(8.15, 6.9)
		}; 
		\addplot[only marks,seagreen,mark size=1.5pt,very thick] coordinates {
			
			(6.95, 2.35)
			(6.86, 3.55)
			(6.7, 4.3)
			(8.13, 4.07)
			(8.53, 2.03)
			
		}; 
		
		\node [right] at (axis cs:    4.13, 3.4) {{\scriptsize{\color{black} UE}}};
		\node [right] at (axis cs:  5, 2.7) {{\scriptsize{\color{black} BS}}};
		
	\end{axis}
\end{tikzpicture}
	\caption[Vor]{A Poisson-Voronoi tessellation model of a cellular network. The BSs, induced by a PPP, partition the plane into Voronoi cells, each having one BS and a number of UEs, with the property that a UE in some cell is closer to the BS of that cell than to the BS of any other cell. This model is so natural and realistic for cellular networks that it has superseded the once popular model of hexagonal cells, which was artificial and idealistic.}
	\label{fig:Voronoi}
\end{figure}
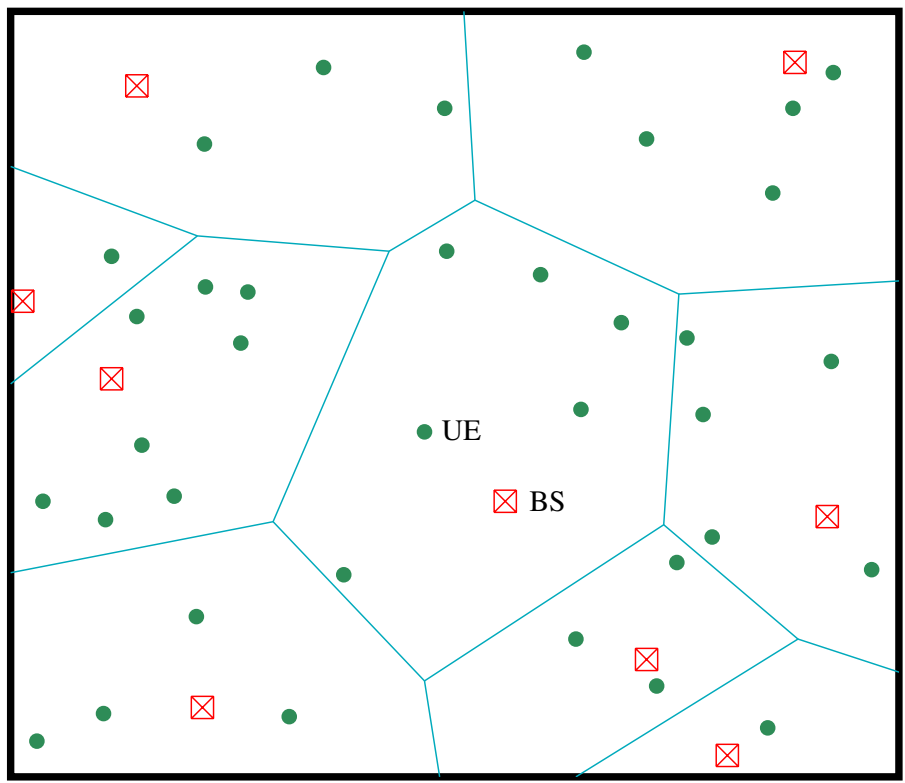

This spatial consideration afforded by stochastic geometry is a key paradigm
shift in communications systems performance evaluation, where time averages
have traditionally reigned supreme. Before the advent of stochastic
geometry, performance analysis of wireless networks was performed using
either damaging mathematical simplifications or exhaustive simulations with
the aim to average out the many sources of randomness, such as BS and user
equipment (UE) locations and fading phenomena. The mathematical
simplifications gave rise to poor analytical results. 

An important technique used in cellular networks, and considered in the
present article, is fractional power control (FPC), where each UE adjusts
its power level in the UL direction under control of its serving BS \cite%
{Haroon20}. More specifically, each UE controls its transmit power such that
the received signal power at its serving BS is equal to a predefined
threshold. This control benefits the UE by saving its energy and benefits
the BS by decreasing its SIR. The optimal levels of transmit power in a
network depend on path loss, shadowing, and multipath fading, as well as the
network configuration. It should be noted, however, that while FPC in
homogeneous cellular networks, it is less effective in heterogeneous
networks (HetNets), where there are more than one tier of BSs, because far
UEs, especially those at the edge of the cell, will use more transmit power,
causing more interference to neighbor cells.

We prove in this article two theorems that expose a prevalent flaw prevalent in the stochastic geometric characterization of coverage
probability, namely that the latter is dependent of BS density. The
Theorems, one for DL and one for UL, prove that the opposite is true; namely
that the coverage probability is independent of BS density. This fact
has only been alluded to in some articles previously \cite{Bai15,Herath18}. The
authors of \cite{Bai15} in the context of analyzing DL coverage probability
of millimeter-wave cellular networks noted that \textquotedblleft coverage
does not scale with BS density." Also, the authors of \cite{Herath18}, while
analyzing UL FPC, noted that coverage is \textquotedblleft \ invariant to the
density of deployment of BSs when the shadowing is mild and power control is
fractional." To the best of our knowledge, the present article is the first
to provide a rigorous mathematical proof that coverage probability is
independent of BS density.

The rest of the article is organized as follows. Section \ref{RW} reviews the
stochastic geometric models of cellular networks in the recent literature. Section \ref{MODEL} provides the
modelling process of a cellular system in both directions, downlink (DL) and
uplink (UL). The simulation algorithms developed to validate the mathematical models are presented in Section \ref{SIM}. Some numerical results for sample systems are presented in Section \ref{EW}. Section \ref{CONC} has the conclusions.

\section{Related work}
\label{RW}
Perhaps the precursor of using stochastic geometry as a modelling vehicle
for wireless systems was the work in \cite{Gilbert61}, approximately half a
century ago. However, the real momentum, especially for cellular networks,
has been gained over the past decade which witnessed an outbreak of hectic
research in the field. This research can be categorized in many ways, but
the direction of transmission, i.e. DL, UL or both, seems suitable enough.

Much work has focused on DL coverage probability. For example, in \cite{Li11}%
, the problem of DL data transmission scheduling in wireless networks is
studied, proposing two algorithms concerning power assignment. In \cite%
{Bai15}, the authors study DL coverage probability in the context of
millimeter-wave (mmWave) cellular networks and note that mmWave coverage
does not scale with BS density. In \cite{Chen19}, DL coverage probability is
studied with composite $\kappa-\mu$ shadowed and lognormal shadowed fading, 
also known as doubly-selective
(DS) fading. In \cite{Qiong20}, the authors derive an expression for DL
coverage probability, taking into account a queue setup. The static
properties of the physical layer of the network are studied by stochastic
geometry and the dynamic properties of the queue setup are studied with a
discrete time Markov chain. Upper and lower bounds on the dynamic coverage
probability are derived. In \cite{Liu20}, the authors study DL in a setting
of an ultra-dense network, a promising technology for 5G cellular networks,
where lots of low power Small Base Stations (SBSs) overlapping with Macro
Base Stations (MBSs) are deployed. The authors derive the coverage
probability.

Many authors have analyzed the DL coverage probability in the context of
HetNets. For example, in \cite{Ouamri20}, the authors characterize the DL
coverage probability in the context of a 2-tier HetNet, under Line of Site
(LOS) and Non Line of Site (NLOS) Path loss models with different path loss
exponents, using stochastic geometry. Also working in HetNets are the
authors of \cite{Fadoul20b}, who consider interference to be a major
performance bottleneck. They consider $K$-tier transmission modeled by
factorial moment and stochastic geometry to obtain expressions for the
coverage probability. They also compare the model with a single-tier,
traditional hexagonal grid model, concluding the superiority of the former.
Also, in \cite{Lei18}, DL is studied using the stochastic geometry for a
two-tier ultradense HetNet with small-cell base stations (SBSs) and UEs
densely deployed in a traditional macrocell network. Performance is studied
in terms of the association probability, average link spectral efficiency
(SE), and average DL throughput. In \cite{Fadoul20}, the authors study DL
for HetNets, where in each typical macrocell, many 
femtocells and few picocells are deployed. They reveal that the coverage probability of the DL direction
decreases when the FPC increases in the UL direction,
mentioning nothing of the deployment density. In \cite{Bouras20}, the authors
note that in current cellular networks, cell association is heavily based on
the DL signal power and that all devices are associated with the same BS in
both DL and UL, noting that this technique is inappropriate in HetNets as
transmission levels significantly vary from BS to BS. They suggest
decoupling DL and UL as a solution, where the UL cell association is not
necessarily based on the same criteria as the DL association. The authors of 
\cite{Arif20} consider also the same decoupling solution to alleviate the
load imbalance problem in HetNets. The BSs in each tier are modeled by an
independent, homogeneous PPP, whereas the spatially clustered UEs are
modeled by a Mat\'{e}rn cluster process. Analytical expressions for the coverage
probability are derived, showing the merit of the solution.

Like DL, a huge body of research work has focused on UL coverage probability
in different scenarios. For example, in \cite{Kouzayha20}, the authors study
UL for a radio frequency (RF) wake-up solution for Internet of Things (IoT)
devices over cellular networks. When the IoT device has no data to transmit,
it turns its main circuitry OFF and switches to a sleep mode. The transition
back to the active mode is only achieved upon receiving enough power at the
device's front end. They derive an expression for the UL coverage
probability after successful wake up. In \cite{Chun20}, the authors study UL
in a densely deployed cellular network in an interference limited scenario,
meaning that the thermal noise power is very small if compared with the
interference. The authors characterize the SIR, noting it is much more
tractable than SINR in a Poisson cellular network. In \cite{Mariam21}, UL is
studied in the context of mmWave cellular networks. The coverage
probability is characterized under the assumption that the BSs
and UEs are modeled as two independent homogeneous PPPs. Interestingly, the
authors find no clear and consistent correlation between the coverage probability and BS
density, noting that the probability increases with increasing BS density at
first, then decreases as the network becomes denser and eventually collapse
as the density is further increased. In \cite{Kamiya20}, the authors employ
stochastic geometry to study the SINR for UL Poisson cellular network, where
truncated fractional transmit power control is performed. They obtains the
UL SINR distribution under channel-adaptive user scheduling, including cases
in which edge users are both allowed and not allowed to transmit at maximum
transmit power. In \cite{Herath18}, the authors analyze three UL transmit
power control schemes, assuming composite Rayleigh-lognormal fading. Using
stochastic geometry, they derive network coverage probability, noting that
it is highly dependent on the severity of shadowing, the power control
scheme, but \emph{invariant} of the density of deployment of base stations
when the shadowing is mild and power control is fractional. This note is
relevant to and consistent with the theorems proved in the present work.

Also, like DL, many authors have studied UL coverage probability for
HetNets. For example, in \cite{Haroon20}, the authors study SIR for HetNets
leveraging FPC. They propose nonuniform SBS deployment (NU-SBSD) to reduce
interference, where SBS deployment (SBSD) near macro base station (MBS) is
avoided, and MBS coverage edge area is enriched with ultra-dense SBSD. Also,
in \cite{Onireti20}, the authors characterize the UL outage probability,
which is the complement of coverage probability, in multitier millimeter
wave cellular networks, using stochastic geometry. In each tier, the BSs are
assumed to have their spatial density, antenna gain, receiver sensitivity,
blockage parameter, and pathloss exponents. The results show that imposing a
maximum power constraint on the user significantly affects the SINR and
outage probability. Moreover, in \cite{Jia19}, the authors indicate that in
the conventionally coupled UL and DL association (CUDA), the UL performance
is limited greatly by the DL parameters such as the density and power of
base stations (BSs). To overcome this issue, they focus on a three-tier
HetNet as well as global performance evaluation, where the cross-tier
dual-connectivity (DC) and decoupled UL and DL association (DUDA) are
integrated. In \cite{Ali19}, UL is studied for a two-tier dense HetNet,
where decoupled UL coverage probability is computed using multi-slope path
loss model. The authors find that the decoupled UL coverage probability is
higher when incorporating multi-slope path loss model as compared to
single-slope path loss model, while the decoupled UL spectral efficiency is
observed to be lower when incorporating dual slope path loss model.

Finally, some research work, such as ours, focuses on both DL and UL,
targeting comprehensive treatment. For example, in \cite{Andrews16}, an
authoritative tutorial on cellular network analysis using stochastic
geometry is provided, characterizing DL and UL coverage probabilities after
a lavish introduction to stochastic geometry. Also, in \cite{Gao19}, both DL
and UL are investigated in the context of mmWave communication, known for
being sensitive to blockage. The authors consider a Time-Division Duplex
(TDD) mode able to provide dynamic UL-DL configurations, and compute the
coverage probability based on stochastic geometry. In \cite{Kundu20}, both
DL and UL are studied in the context of full duplex (FD) for a UE capable of
transmitting and receiving data simultaneously in the same frequency
resource. The authors obtain the coverage probabilities, revealing that
while FD improves DL performance, it severely hurts UL performance.
Furthermore, in \cite{Sadeghabadi20}, the authors study both DL and UL for
massive multiple-input multiple-output (M-MIMO), promising for increasing
the spectral efficiency. They analyze an asynchronous DL M-MIMO system in
terms of the coverage probability by means of stochastic geometry.

Some authors investigate coverage probability for both DL and UL, but in the
context of HetNets. An interesting example is in \cite{Wang19} where the
authors consider DL simultaneous information transmission and power transfer
and UL information transmission for an unmanned aerial vehicle
(UAV)-assisted mmWave cellular network. Distinguishing features of mmWave
communications, such as different path loss models and directional
transmissions are taken into account. DL association probability and energy
coverage of the two tiers of UAVs and ground base stations (GBSs) are
investigated.

Still, some other authors study both DL and UL for the purpose of
decoupling, aiming to alleviate BS load imbalance and also ease
interference. For example, in \cite{Bouras20}, the authors note that in
current cellular networks, cell association is heavily based on the DL
signal power and that all devices are associated with the same BS in both DL
and UL. While this practice is adequate in homogeneous networks, where all
BSs have similar transmission levels, it can fail in dense HetNets as
transmission levels significantly vary from BS to BS. They propose
decoupling DL and UL as a solution, where the UL cell association is not
necessarily based on the same criteria as the DL association. Also, the
authors of \cite{Arif20} consider DL and UL decoupling is investigated to
alleviate the load imbalance problem in HetNets, aiming to increase the
coverage probabilities and data rates. The BSs in each tier are modeled by
an independent, homogeneous PPP, whereas the spatially clustered UEs are
modeled by a Mat\'{e}rn process. Analytical expressions for the coverage
probability, are derived.

\section{modelling}
\label{MODEL}
The subject of stochastic geometry and its use in modelling wireless systems
is vast \cite{Blaszczyszyn18}. The key aspect of the present study is that
all the BSs are located according to a PPP $\Phi $, which effectively means
they are randomly scattered in the Euclidean plane with independent
locations. At any given time, only one UE can be active communicating with
the BS on any time/frequency resource. Before delving even more into the
modelling process, some definitions used throughout the article are in order.

\begin{description}
	\item \textbf{Definition 1 (BS-UE association)}: BS-UE association is the assignment of a UE to a BS, for both to
	establish a communications session. 
	
	\item \textbf{Definition 2 (Serving BS)}: Once a UE is associated with a BS, the latter is said to be the serving BS of the UE.
	
	\item \textbf{Definition 3 (Typical receiver)}: The typical receiver is the receiving device (UE or BS) where the SIR is to be assessed. It is always placed at the origin of the Euclidean plane in the
	model, or the origin of the simulation window in the simulation.
	
	\item \textbf{Definition 4 (Tagged transmitter)}: The tagged transmitter is transmitting device (UE or BS)  associated with the typical receiver.
	
	\item \textbf{Definition 5 (Typical circle)}: The typical circle is the circle centered at the typical
	receiver and having the tagged transmitter on its circumference.
	
	\item \textbf{Definition 6 (Signal)}: A signal is the transmission arriving at, and intended for,
	the typical receiver.
	
	\item \textbf{Definition 7 (Interference)}: An interference is the transmission arriving at, but not intended for, the typical
	receiver.
	
	\item \textbf{Definition 8 (Interferer)}: An interferer is a transmitter causing interference at the typical receiver. That is, it is any transmitter in the network other than
	the tagged transmitter.
	
	\item \textbf{Definition 9 (Signal to Interference Ratio (SIR))}: The quotient of
	the signal at the typical receiver and the sum of all 
	interferences at the typical receiver.
	
	\item \textbf{Association rule}: A UE will associate with the BS nearest it.
\end{description}
As per Definitions 3 and 4, in DL the typical receiver is a UE and the
tagged transmitter is a BS, whereas in UL the typical receiver is a BS and
the tagged transmitter is a UE. 

As per the association rule, a UE will be nearer to its serving BS than to
any other BS in the cellular network. Strictly speaking, however, the UE
associates with the BS that provides the highest average SIR. Surprisingly,
depending on the fading and shadowing conditions, a more distant BS could
provide a higher instantaneous SIR, hence gets associated with the UE. That
is why, we rely on the average, which invariably translates to association
with the nearest BS. We will denote the distance between the associated
BS-UE pair throughout by $R$. If the BS density is $\lambda $, it can be
shown that $R$ is a random variable (RV) with the Rayleigh distribution\ 
\begin{equation}
	f_{R}(r)=2\lambda \pi re^{-\lambda \pi r^{2}},\qquad r\geq 0
	\label{Reighlay}
\end{equation}

Along the same line, the notation used throughout the article is provided in
Table \ref{table:Notation}. The notation of the PPP, pivotal in our work,
depends on its setting and the field it is being applied in and on the
interpretation of the process. For example, a simple PPP $\Phi $ may be
considered as a random set, which suggests the notation $x\in \Phi $,
implying that $x$ is a random point belonging to or being an element of the
Poisson random set $N$. Another, more general, interpretation is to consider
a Poisson or any other point process as a random counting measure, so one
can write the number of points of a PPP $\Phi $ being found or located in
some (Borel measurable) region $B$ as $\Phi (B)$, which is an RV (having a
Poisson distribution, in the case of a PPP) \cite{Blaszczyszyn18}.

\begin{table}[H]
	\caption{Notation used in the model and simulation.} 
	\centering 
	\begin{tabular}{c l } 
		\hline\hline 
		Parameter & Description\\ [0.5ex] 
		\hline 
		BS & Base station\\
		UE & User Equipment (can be a mobile phone, tablet, laptop, etc.)\\
		$\Phi$ & Poisson point process (PPP) of BSs\\
		$\Psi$ & Point process of UEs (not Poisson) \\
		$\lambda$ & Density of BS (per m$^2$), i.e. intensity of PPP $\Phi$\\  
		$\alpha$ & Path-loss exponent (per m) \\
		SIR & Signal to interference ratio (dB) \\
		$\xi$ & SIR threshold (dB)\\
		$G$ & Rayleigh channel gain of tagged transmitter ($G \sim Exp(1)$)\\ 
		p & Transmit power (Watts)\\
		$p_d$ & DL coverage probability\\
		$p_u$ & UL coverage probability\\
		\hline
		$\mathcal{N}$ & Number of simulation runs (PPP realizations)\\
		$S$ & Side of the simulation square window (m)\\
		$N$ & Number of BSs in a simulation run, $N \sim Pois(\lambda S^2)$).\\
		($x,y$) & Location of a BS in a simulation run\\
		($u,v$) & Location of a UE in a simulation run\\
		Covered & No. of times typical receiver is covered by tagged transmitter\\
		\hline 
	\end{tabular}
	\label{table:Notation} 
\end{table}

\begin{figure}[H]
	\centering
	\begin{subfigure}{0.2\textwidth}
		\caption{ \vspace{-1cm} \hspace{-10cm}\small (a)}
		\begin{tikzpicture}[scale=1.5]
			\hspace{-3cm}
			\begin{axis}[,
				height=8.5cm,
				width=8cm,
				hide axis,
				legend style={at={(0.55,-.01)}, anchor=north,legend columns=-3,mark options={scale=0.75}},
				]
				
				\definecolor{sienna}{rgb}{0.53, 0.18, 0.09}	
				
				\addplot [only marks, mark size=2.5pt, thin, color=red, mark=BS,mark options={scale=1}] 
				coordinates {
					(0,0)
					(1.5,4.14)
					(-1.7,3.59)
					(-0.46,2.71)
					(2.5,3.22)
					(-2.4,1.28)
					(-0.24,1.30)
					(1.05,1.53)
					(3.3,1.86)
					(3,0.36)
					(-2.6,-0.62)
					(-0.7,-0.28)
					(3,-1)
					(-0.57,-1.5)
					(1.7,-1.39)
					
				};
				
				\addplot [only marks,  mark size=1.5,color=seagreen,very thick,mark options={scale=1}] coordinates {
					(-1.05,4.54)
					(2.3,4.54)
					(0.08,3.69)
					(2.15,3)
					(-1.64,1.02)
					(-0.31,0.65)
					(1.30,0.80)
					(3.07,1.28)
					(1.22,0.10)
					(2.18,0.07)
					(-0.28,-0.56)
					(3.87,-0.48)
					(-3.44,-1.24)
					(-0.06,-1.17)
					(2.37,-1.61)
				};
				
				\addplot [only marks,  mark size=5,color=black, mark=+] coordinates {
					(1.22,0.10)
				}; 

				\draw[cyan, thick=1pt] (axis cs:1.22,0.10) circle (0.9cm);
				\draw[blue,thick=1pt] (axis cs:0,0) -- node[centered, above=-1mm,black]{\scriptsize$R$} (axis cs:1.22,0.10);
				\draw[dashed,thick=1pt,seagreen] (axis cs:1.22,0.10) -- node[centered, above right=-2.3mm,black]{\scriptsize $D_z$} (axis cs:-0.49,2.71);
				\draw[red,thick=1pt] (axis cs:0.08,3.69) -- node[centered, above left= -2mm,black]{\scriptsize $R_z$} (axis cs:-0.46,2.71);
				
				\put(105,202){\shortstack{$z$}}
				
			\end{axis}   
		\end{tikzpicture}
		\label{fig:DL}
	\end{subfigure}

	\rule{.7\textwidth}{0.01in}
	\vspace{0.6cm}
	
	\hspace{-0.35cm}

	\begin{subfigure}{0.2\textwidth}
		\vspace{-1.2cm}
		\caption{ \vspace{-1cm} \hspace{-10cm}\small (b)}
		\begin{tikzpicture}[scale=1.5]
			\hspace{-3cm}
			\pgfplotsset{
				every axis legend/.append style={ at={(1.05,0.95)}, anchor=north west,legend columns = 1}}
			\begin{axis}[,
				height=8.5cm,
				width=8cm,
				hide axis,
				]
				
				\addplot [only marks, mark size=2.5pt, thin, color=red, mark=BS] 
				coordinates {
					
					(0,0)
					(1.5,4.14)
					(-1.7,3.59)
					(-0.46,2.71)
					(2.5,3.22)
					(-2.4,1.28)
					(-0.24,1.30)
					(1.05,1.53)
					(3.3,1.86)
					(3,0.36)
					(-2.6,-0.62)
					(-0.7,-0.28)
					(3,-1)
					(-0.57,-1.5)
					(1.7,-1.39)
					
				};
				
				\addplot [only marks,  mark size=1.5,color=seagreen, very thick] coordinates {
					
					(-1.05,4.54)
					(2.3,4.54)
					(0.08,3.69)
					(2.15,3)
					(-1.64,1.02)
					(-0.31,0.65)
					(1.30,0.80)
					(3.07,1.28)
					(1.22,0.10)
					(2.18,0.07)
					(-0.28,-0.56)
					(3.87,-0.48)
					(-3.44,-1.24)
					(-0.06,-1.17)
					(2.37,-1.61)
				};
				\addplot [only marks,  mark size=5,color=black, mark=+] coordinates {
					(0,0)
				};
				\definecolor{sienna}{rgb}{0.53, 0.18, 0.09}	
				
				\draw[cyan, thick=1pt] (axis cs:0,0) circle (0.9cm);
				\draw[ thick=1pt, blue] (axis cs:0,0) -- node[centered, above= -1mm,black]{\scriptsize$R$} (axis cs:1.22,0.10);
				\draw[ thick=1pt, dashed, seagreen] (axis cs:0,0) -- node[centered, above right= -1mm,black]{\scriptsize $U_{\mathfrak{z}}$} (axis cs:0.08,3.69);
				\draw[thick=1pt, red] (axis cs:0.08,3.69) -- node[centered, above left= -2mm,black]{\scriptsize $R_{\mathfrak{z}}$} (axis cs:-0.46,2.71
				);
				\put(138,240){\shortstack{$\mathfrak{z}$}}
				
			\end{axis}   
		\end{tikzpicture}
		\label{fig:UL} 
	\end{subfigure}
	
	\caption[T]{Stochastic geometric models for assessing the SIR at a typical receiver at the origin of the cellular network. (a) The DL model, where the typical receiver is a UE. The typical circle defines an exclusion zone, as it cannot contain a BS inside. The BSs outside the typical circle, which cause interference at the typical UE, form a PPP $\Phi$. (b) The UL model, where the typical receiver is a BS. The typical circle does not define an exclusion zone, as it can contain UEs inside. All the UEs form a PP $\Psi$. Since the UEs are {\emph satellites} to their serving BSs, as per the association rule, $\Psi$ is not Poisson, a major challenge to the analysis. We mitigate this challenge by relocating each UE, except the tagged, to the position of its serving BS. That is, we relocate all the interferers.
	} 
		\label{fig:DU} 
\end{figure}
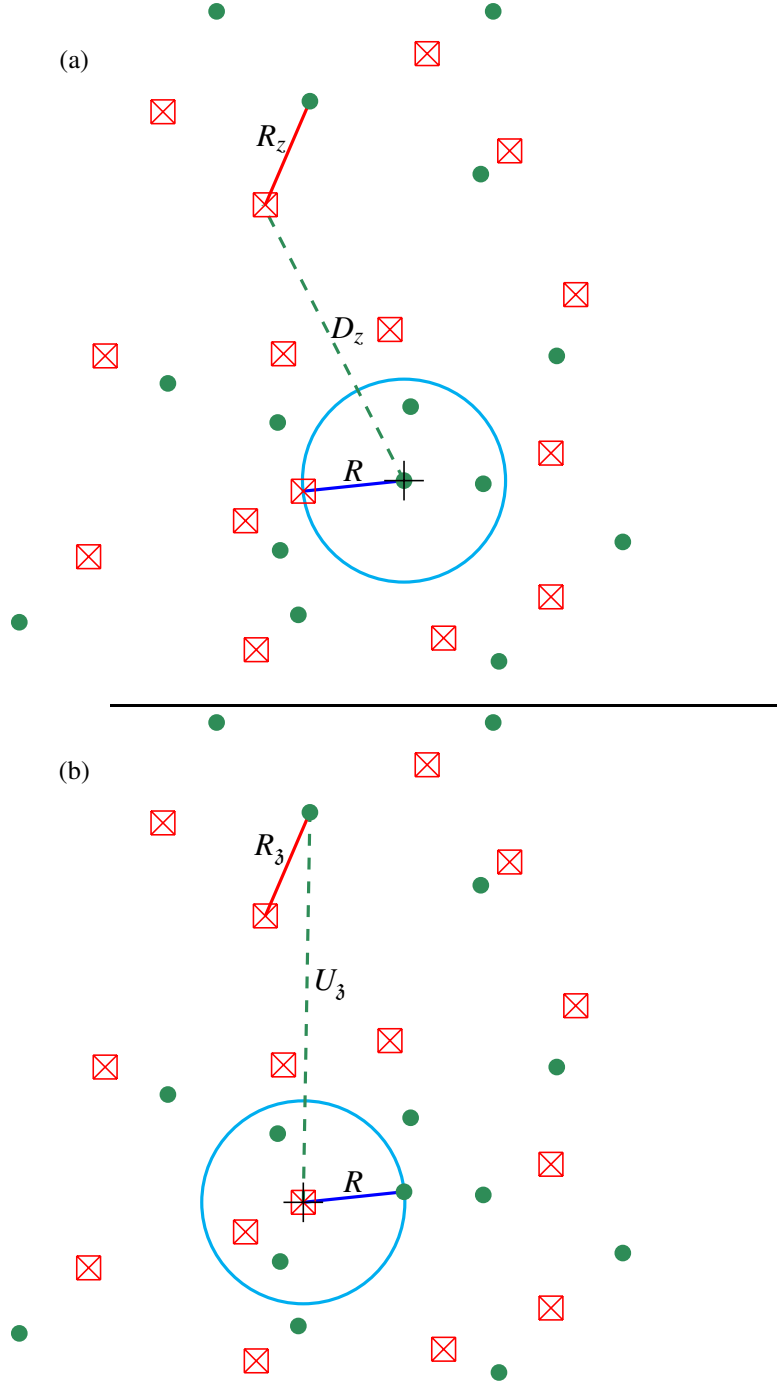

Figure \ref{fig:DU} shows a stochastic geometric construction to
characterize the SIR at the typical receiver of a cellular network, which with the tagged transmitter, 
defines the typical circle. In part \ref{fig:DL}, we can see
the DL model, where the typical receiver is a UE, and the tagged transmitter
is a BS at distance $R$. All the BSs outside the typical circle are
interferers to the typical UE. The typical circle defines an exclusion zone,
in the sense that there can be no BS inside of it (or else the typical UE
would associate with it.) In part \ref{fig:UL}, we can see the UL model, where the typical
receiver is a BS, and the tagged transmitter is a UE at distance $R$. All
the UEs except the tagged are interferers to the typical BS. The
typical circle does not define an exclusion zone, as there can be UEs inside
of it as we can see (the association problem of the DL model is not present here.)

The RV $R_{z}$ are identically distributed but not independent in general.
The dependence is induced by the structure of Poisson-Voronoi tessellation
and the restriction that only one BS can lie in each cell. To visualize this
dependence, recall that the presence of a BS in a particular Voronoi cell
forbids the presence of any other BS in that cell. However, as discussed in
detail in the next section, this dependence is weak, motivating the
approximation we will make in the UL model below. On the other hand, the RV $%
R_{z}$ is upper bounded by the RV $D_{z}$, even for the pairs inside the
typical circle (or else the BS at $z$ would be closer to the typical UE,
violating the association rule.). The UEs are satellites to the points of
the PPP $\Phi $, based on the association rule, so they are not themselves a
PPP, making the UL model a challenge. This challenge is mitigated in the
present work through an elegant approximation. 

We can relate Figure \ref{fig:Voronoi} and Figure \ref{fig:DU} by
recognizing that in the present article, we consider orthogonal
communications. This means that in each cell there can be only one active UE
on any time/frequency resource. Accordingly, Figure \ref{fig:Voronoi} is a
snapshot of the UEs that are active in all the cells at the same time. Thus,
we see a random number of UEs in each cell. Figure \ref{fig:DU}, on the
other hand, is a snapshot of the UEs that are active on the same frequency
in all the cells at the same time. Thus, we see exactly one UE per cell. We
focus in the model on one such resource. That is, every BS-UE associated
pair in Figure \ref{fig:DU} is operating on the same resource, hence is the
interference that we are going to characterize.

Given a homogeneous PPP with intensity $\lambda $, the number of process
points in the an area of size $A$, denoted by $\Phi (A)$, is Poisson
distributed%
\begin{equation*}
	\mathbb{P}[\Phi (A)=k]=\frac{e^{-\lambda |A|}(\lambda |A|)^{k}}{k!}
\end{equation*}%
That is, a homogeneous PPP is completely characterized by a single number, $%
\lambda $. It is both stationary (translation invariant) and isotropic
(rotation invariant).

Similarly, if $F$ is a continuous RV, we can also write its Laplace transform as a
(continuous) expectation, namely
\begin{equation}
	\mathcal{L}_{F}(s)=\mathbb{E}\left[ e^{-sF}\right]   \label{LapAsExpect}
\end{equation}

The probability generating function of a non-negative integer-valued RV
leads to the probability generating functional (PGFL) with respect to any
non-negative bounded function $f$ for a point process defined on $\mathbb{R}%
^{d}$, with $0\leq f(x)\leq 1$. The need for the PGFL, which we will invoke
twice in this article, arises naturally when one wishes to find the Laplace
transform of a sum of functions. As the Laplace transform of a RV is already
an expectation of an exponential function, with the RV its exponent, then if
the RV is a sum of independent and identically distributed (iid) RVs, the
exponential becomes a product lending itself readily to the invocation of
PGFL. Due to this intimate relationship, some authors call the PGFL in this
context a Laplace functional.

We will also use in the present work Slivnyak's theorem \cite{Blaszczyszyn18}%
. It states that for a PPP, because of the independence between its points,
conditioning on a point at some location $x$ does not change the
distribution of the rest of the process. That is, removing an
infinitesimally small area from the underlying plane of the PPP, leaves
still a PPP, since the distribution of points in all nonoverlapping areas
are independent for the PPP. Consequently, any property seen from some
location $x$ remains the same whether or not there is a point at $x$. Though
simple, this theorem allows us to add and  remove points to/from the PPP
liberally. In our work, for example, it allows us in UL to place a BS from
those imposed by the PPP $\Phi $\ and place it at the origin, treating the
remaining points, $\Phi \backslash x$, still as a PPP (for example, when
invoking a PGFL.) 

Before starting the modelling process, a word on random channels is in
order. In our work, random channel effects are incorporated by
multiplicative RVs, namely $G$ for the signal and $G_{i}$ for each
interferer $i$. For simplicity we assume these all correspond to Rayleigh
fading with mean $1$, denoted $\exp (1)$, i.e. $\mu =1$ in $f_{G_{i}}(x)=\mu
e^{-\mu x}$. We consider small-scale Rayleigh fading, and assume $G$ and the 
$G_{i}$ iid RVs following an exponential distribution with mean $1$.

\subsection{Downlink model}

The key assumptions of the DL system model are as follows:

\begin{itemize}
	\item BSs are located according to a homogeneous PPP $\Phi $ of intensity $%
	\lambda $ in the Euclidean plane. 
	
	\item A UE wishing to start a communications session associates with the BS
	that is closer to it than any other BS in the cellular network (association
	rule.)
	
	\item The BS transmits at a fixed power $p$ to a designated UE on a
	particular time-frequency resource, i.e. orthogonal multiple access
	communications within the cell take place. The consequence of orthogonality
	is that the UE sees interference from all other BSs in the plane using the
	same resource.
	
	\item Signals attenuate with distance according to the standard power-law
	path loss propagation model, with path loss exponent $\alpha >2$. That is,
	the average received power at distance $r$ from a transmitter of power $p$
	is $pr^{-\alpha }$.
	
	\item Random channel effects are incorporated by a multiplicative RVs $G$
	for the signal and $G_{z}$ for every interferer at $z$. For simplicity we
	assume these all to correspond to Rayleigh fading with mean $1$, so $G$ and
	the $G_{z}$ are iid RVs, having exponential distribution with mean $1$.
\end{itemize}

Let $I_{d}$ denote the interference experienced at the typical UE. The interference is due to every BS $z$ in the plane, except the tagged BS, denoted by $\mathfrak{b}$, at
distance $D_{z}$ from the typical UE, as shown in Figure \ref{fig:DL},.  That is, the interference $I_{d}$ is created
by a PPP with intensity $\lambda$ outside the typical circle, and is given by
\begin{equation}
	I_{d}=\sum_{z\in \Phi \backslash \{ \mathfrak{b}\}}pG_{z}D_{z}^{-\alpha},
	\label{id}
\end{equation}%
where $p$ is the power of the BS at point $z$, and $D_{z}$ is a RV representing the
distance from the BS at $z$a and the typical UE. The
SIR at the typical UE is then given by
\begin{equation}
	\mathtt{SIR}_{\text{UE}}=\frac{pGR^{-\alpha }}{I_{d}}  \label{SIRd}
\end{equation}%

Note that the distance $R_{z}$ is not used in anything in DL, but it will be used
in the UL model. Note also that the typical circle defines an exclusion zone. That is, there can never be a BS closer to the
typical UE than the tagged BS. For if there were such a
BS, it would be the tagged BS associated with the typical UE. Later in the UL
model, we will see that we cannot define an exclusion zone by any BS-UE
pair. 

The goal now is to derive the DL coverage probability $p_{d}$ in the
cellular network, which is exactly the complementary cumulative distribution function (CCDF) of SIR over the entire
network (Recall that the CDF gives $\mathbb{P}[$SIR$\leq \xi ]$). The
coverage probability can also be visualized as the probability that a randomly
chosen user can achieve a target SIR $\xi $, the average fraction of users
who at any time achieve SIR $\xi $, or the average fraction of network area
that is in \textquotedblleft coverage\textquotedblright \ at any time. 
We
have two routes now for obtaining $p_{d}$ as far as the manipulation of RVs is
concerned. Either use RVs throughout and keep resolving them via the $%
\mathbb{E}$ operator vehicle, or condition the SIR expression at the very
beginning, using particular values for those RVs, then keep on
deconditioning. We will take the former route. At the outset, We will invoke the concept of total probability, using (\ref{id}) and (\ref{SIRd}), to get
\begin{eqnarray}
	p_{d|R} &=&\mathbb{P}[\mathtt{SIR}_{\text{UE}}>\xi ]  \notag \\
	&=&\mathbb{P}\left[ \frac{pGR^{-\alpha }}{I_{d}}>\xi \right]   \notag \\
	&=&\mathbb{P}\left[ G>\frac{\xi }{p}R^{\alpha }I_{d}\right]   \notag \\
	\overset{(\text{a})}{=} &&\mathbb{E}%
	_{I_{d}}\left[ \mathbb{P}\left[ G>\frac{\xi }{p}R^{\alpha }I_{d}\right] %
	\right]   \notag \\
	\overset{(\text{b})}{=} &&\mathbb{E}%
	_{I_{d}}\left[ e^{-\frac{\xi }{p}R^{\alpha }I_{d}}\right]   \notag \\
	\overset{(\text{c})}{=} &&\mathcal{L%
	}_{I_{d}}(\frac{\xi }{p}R^{\alpha })  \label{First}
\end{eqnarray}%
where%
\begin{equation}
	\mathcal{L}_{A}(s)=\int_{0}^{\infty }e^{-st}f_{A}(t)dt=\mathbb{E}\left[
	e^{-sA}\right]   \label{LT}
\end{equation}%
is the Laplace transform of the RV $I_{d}$\ conditioned on the RV $R$\ between the typical UE and the tagged BS. In (a) we utilized the
fact that we can write a probability $\mathbb{P}\left[ A>B)\right] $ as $%
\mathbb{E}_{B}\left[ \mathbb{P}\left[ A>B\right] \right] $ (or $\mathbb{E}%
_{A}\left[ \mathbb{P}\left[ A>B\right] \right] $), in (b) we benefited from
the fact that $G\sim \exp (1)$, i.e. $f_{G}(r)=e^{-r}$, and in (c) we used
the Laplace transform definition (\ref{LT}).

Clearly, the DL coverage probability $p_{d}$ in (\ref{First}) is conditioned
on $R$, the distance between the typical UE and the tagged BS. We will now embark on deconditioning $p_{d}$ (basically by obtaining its
expectation with respect to $R$.) Since $R$ is the distance between the typical UE
and the closest BS (the tagged BS), it is Rayleigh distributed, i.e. and $%
f_{R}(r)=2\lambda \pi re^{-\lambda \pi r^{2}}$ from (\ref{Reighlay}).
Further, $R$ ranges from an arbitrarily small positive real number greater
than $0$ (to exclude the typical UE) to $\infty $. Thus, the
conditional coverage probability 
\begin{eqnarray}
	p_{d|R} &=&\mathbb{E}_{R}\left[ p_{d|R}\right]   \notag \\
	&=&\mathbb{E}_{R}\left[ \mathcal{L}_{I_{d}}(\frac{\xi }{p}R^{\alpha })\right]
	\notag \\
	&=&\int_{0}^{\infty }\mathcal{L}_{I_{d}}(\frac{\xi }{p}r^{\alpha })f_{R}(r)dr
	\notag \\
	&=&2\lambda \pi \int_{0}^{\infty }e^{-\lambda \pi r^{2}}\mathcal{L}_{I_{d}}(%
	\frac{\xi }{p}r^{\alpha })rdr  \label{pcDecond}
\end{eqnarray}

Next, we will embark on finding the Laplace transform $\mathcal{L}_{I_{d}}$
of the DL interference $I_{d}$. Using (\ref{id}) and (\ref{LT}), we get%
\begin{eqnarray}
	\mathcal{L}_{I_{d}}(s) &=&\mathbb{E}\left[ e^{-sI_{d}}\right]  \notag\\
	&=&\mathbb{E}_{\Phi ,G_{z}}\left[ e^{-s\sum_{z\in \Phi \backslash \{%
			\mathfrak{b}\}}pG_{z}D_{z}^{-\alpha }}\right]   \notag\\
	&=&\mathbb{E}_{\Phi ,G_{z}}\left[ \prod \limits_{z\in \Phi \backslash \{%
		\mathfrak{b}\}}e^{-spG_{z}D_{z}^{-\alpha }}\right]   \notag\\
	\overset{(\text{a})}{=} &&\mathbb{E}_{\Phi }\left[ \prod \limits_{z\in \Phi
		\backslash \{ \mathfrak{b}\}}\mathbb{E}_{G_{z}}\left[ e^{-spG_{z}D_{z}^{-%
			\alpha }}\right] \right]  \notag\\
	\overset{(\text{b})}{=} &&\mathbb{E}_{\Phi }\left[ \prod \limits_{z\in \Phi
		\backslash \{ \mathfrak{b}\}}\mathcal{L}_{G_{z}}(spD_{z}^{-\alpha })\right] 
	 \notag\\
	\overset{(\text{c})}{=} &&\exp \left( -\lambda \int_{\mathbb{R}%
		^{2}\backslash D(o,r)}(1-\mathcal{L}_{G_{z}}(spD_{z}^{-\alpha }))\right) 
	\label{LIds}
\end{eqnarray}%
where $D(o,r)$ is a disc centered at the origin and has a radius $r$. In (a)
we benefited from the independence of the $G_{z}$, which are iid and in (b)
we used the definition (\ref{LT}) of the Laplace transform. In (c), to
decondition on $D_{z}$ which is distributed \emph{differently} for each
point $z$ of the PPP, we invoked the PGFL $\mathbb{E}_{\Phi }\left[
\prod \limits_{z\in \Phi }f(x)\right] $, with $f(x)=\mathcal{L}%
_{G_{z}}(spx^{-\alpha })$, of the PPP $\Phi $. Note that $\mathbb{E}_{G_{z}}$
will be taken at the level of a single point $z$, whereas $\mathbb{E}_{\Phi }
$ will be taken at the level of the PPP $\Phi $. However, for the \emph{%
	expectation with respect to} $\Phi $, we will apply it to $D_{z}^{-\alpha }$%
, as the latter involves all the PPP points. Note finally that some authors
call the PGFL the Laplace functional, when the former is invoked to find a
Laplace transform as is the case here.

Note that $D_{z}$ ranges from $r^{+}$ (to exclude the tagged BS) to $\infty $. That is, the integral is carried out outside the
typical circle, namely from $r$ to $\infty $. Switching to
polar coordinates, with the interferer now at $(x,\theta )\in \mathbb{R}^{2}$, then using the fact that $G_{z}\sim \exp (1)$, i.e. $f_{G_{z}}(t)=e^{-t}$, then Using (\ref{LIds}) yields
\begin{eqnarray}
	\mathcal{L}_{I_{d}}(s) &=&\exp \left( -\lambda \int_{0}^{2\pi
	}\int_{r}^{\infty }(1-\mathcal{L}_{G_{z}}(spx^{-\alpha }))xdxd\theta \right) 
	\text{ }  \notag \\
	&=&\exp \left( -2\pi \lambda \int_{r}^{\infty }(\frac{spx^{-\alpha }}{%
		1+spx^{-\alpha }})xdx\right)   \label{Lap3}
\end{eqnarray}%
For use in (\ref{pcDecond}), we write this result as%
\begin{eqnarray*}
	\mathcal{L}_{I_{d}}(\frac{\xi }{p}r^{\alpha }) &=&\exp \left( -2\pi \lambda
	\int_{r}^{\infty }(\frac{(\frac{\xi }{p}r^{\alpha })px^{-\alpha }}{1+(\frac{%
			\xi }{p}r^{\alpha })px^{-\alpha }})xdx\right)  \\
	&=&\exp \left( -\pi \lambda r^{2}\xi ^{\frac{2}{\alpha }}\int_{\xi
		^{-2/\alpha }}^{\infty }\frac{1}{1+u^{\alpha /2}}du\right) 
\end{eqnarray*}%
where $u=\left( x/r\right) ^{2}\xi ^{-\frac{2}{\alpha }}$. Substituting this
in (\ref{pcDecond}), we get

\begin{equation}
	p_{d}=2\widetilde{\lambda }\int_{0}^{\infty }e^{-\widetilde{\lambda }%
		r^{2}}e^{-\widetilde{\lambda }r^{2}\sqrt[\kappa ]{\xi }\int_{\frac{1}{\sqrt[%
				\kappa ]{\xi }}}^{\infty }\frac{1}{1+u^{\kappa }}du}rdr  \label{pdWithLampda}
\end{equation}%
where $\widetilde{\lambda }=\lambda \pi $ and $\kappa =\alpha /2$.

From (\ref{pdWithLampda}), it \emph{appears} that the DL coverage probability $p_{d}$ is dependent on the BS density parameter $\lambda $. Indeed, this parameter exists in a plethora of coverage probability (and derivatives) expressions all over the literature on stochastic geometric modelling of DL cellular systems (See, for example, 
\cite{ElSawy17, Chun20, Liu20, Kouzayha21, Bai15, Chen19, Qiong20, Ouamri20, Lei18, Fadoul20, Gao19, Ali19, Andrews16, Kundu20, Sadeghabadi20, Wang19}). However, this dependence is false, as we will show in the next
Theorem.

\begin{description}
	\item \textbf{Theorem 1}: Under the stochastic geometric model of the cellular DL
	system, the DL coverage probability $p_{d}$ is independent of the BS density 
	$\lambda $.
\end{description}

\textbf{Proof}: The proof is attained through two changes of
variables. Starting with (\ref{pdWithLampda}), use the substitution $x=r^{2}$
to get

\begin{eqnarray*}
	p_{d} &=&2\widetilde{\lambda }\int_{0}^{\infty }e^{-\widetilde{\lambda }%
		r^{2}}e^{-\widetilde{\lambda }r^{2}\xi ^{\frac{2}{\alpha }}\int_{\xi ^{-%
				\frac{2}{\alpha }}}^{\infty }\frac{1}{1+u^{\kappa }}du}rdr \\
	&=&\widetilde{\lambda }\int_{0}^{\infty }e^{-\widetilde{\lambda }x\left( 1+%
		\sqrt[\kappa ]{\xi }\int_{\frac{1}{\sqrt[\kappa ]{\xi }}}^{\infty }\frac{1}{%
			1+u^{\kappa }}du\right) }dx
\end{eqnarray*}%
Now use the substitution $z=\widetilde{\lambda }x$ to get

\begin{eqnarray}
	p_{d} &=&\int_{0}^{\infty }e^{-z\left( 1+\sqrt[\kappa ]{\xi }\int_{\frac{1}{%
				\sqrt[\kappa ]{\xi }}}^{\infty }\frac{1}{1+u^{\kappa }}du\right) }dz 
	\notag \\
	&=&\frac{1}{1+\sqrt[\kappa ]{\xi }\int_{\frac{1}{\sqrt[\kappa ]{\xi }}%
		}^{\infty }\frac{1}{1+u^{\kappa }}du}
	\label{pdWithoutLampda}
\end{eqnarray}%
where $\lambda $ has totally disappeared, proving the theorem. $\blacksquare$

Theorem 1 then calls for a revisit to the DL coverage probability expressions (and their derivatives)
that include $\lambda $. Those expressions exist in a huge number of articles, such as those cited just above the Theorem. It is worth examining them for possible elimination of a parameter that may exist superfluously. 

\textbf{DL Special cases}:

First, we will consider the special case of $\alpha =4$, i.e. $\kappa =2$. From (\ref{pdWithoutLampda}), we can obtain (using a computer algebra
system, e.g. Maxima (R)), a simple closed form expression for the DL coverage
probability, namely%
\begin{eqnarray}
	p_{d} &=&\frac{1}{1+\sqrt{\xi }\int_{\frac{1}{\sqrt{\xi }}}^{\infty }\frac{1%
		}{1+u^{2}}du} \notag\\
	p_{d} &=&\frac{1}{1+\sqrt{\xi }\left( \frac{\pi }{2}-\arctan (\frac{1}{\sqrt{%
				\xi }})\right) }
	\label{pd4}
\end{eqnarray}%

Second, we will consider the special case of $\alpha =6$, i.e. $\kappa =3$. From (\ref{pdWithoutLampda}), we can obtain (using a computer algebra
system, e.g. Maxima (R)), a not-so-simple closed form expression for the DL coverage
probability, namely%
\begin{eqnarray}
	p_{d} &=&\frac{1}{1+\sqrt[3]{\xi }\int_{\frac{1}{\sqrt[3]{\xi }}}^{\infty }%
		\frac{1}{1+u^{3}}du} \notag\\
	p_{d} &=&\frac{6c}{6c+\ln \left( c^{2}-c+1\right) +\sqrt{3}\left( {\pi }%
		-2\arctan \left( \frac{2c-1}{\sqrt{3}}\right) \right) -2\ln \left(
		c+1\right) }
	\label{pd6}
\end{eqnarray}%
where $c=1/\sqrt[3]{\xi }$. To further attest to the accuracy of the last
two results, they will be plotted later together with the corresponding
simulation counterparts.

\subsection{Uplink model}

The net interference at the typical BS is the sum of the received transmissions 
from all the UEs except the tagged, including those inside the typical circle. Referring to Figure \ref{fig:UL}, for each UE $\mathfrak{z}\in \Psi $, we denote its distance to its serving BS by $%
R_{\mathfrak{z}}$. Although the RVs $R_{\mathfrak{z}}$ are identically
distributed, they are not independent. However, as shown later in this
section, this dependence is weak and we will henceforth assume the $R_{\mathfrak{z}}$ iid. Under this assumption, we will first derive the coverage probability for the general distribution of $R_{\mathfrak{z}}$.

The set of interferers are the points of $\Psi $, which is not a PPP. The reason is
that they are not scattered uniformly in $R^{2}$, but rather associated to a
PPP as satellites. This poses difficulty, as we will not have the luxury of
using such useful tools as the PGFL here. However, we can get around this
difficutly as follows. Note that each point of $\Psi $ is associated, by
being closest, to a point in the PPP $\Phi $ of BSs, which we used above in
the down link analysis. More importantly, as each point in $\Psi $ is
located somewhere around the corresponding point in $\Phi $, we can
approximate the "spatial" average of the former to be the latter.
Consequently, we can approximate the locations of the interfering UEs $\Psi $
by the points of $\Phi $. Accordingly, below we will carry out the sums and
products involved in the stochastic geometry analysis over $\Phi $, rather
than $\Psi $. Specifically, for the sake of calculating the interference, we
will consider that each interfering UE is placed exactly at
its serving BS's location. Consequently, referring to Figure \ref{fig:UL},
we will employ the distance $R_{\mathfrak{z}}$ between this UE and its
serving BS to calculate its emitted power. We will then consider this as
interference at the typical BS at distance $D_{z}$ away, not $U_{\mathfrak{z}}$. 

In UL, FPC leads to amplifying the transmit power $p$ at the UE based on its
distance to the serving BS. If the distance is $R$ and the FPC factor is $%
\epsilon $, with values in $[0,1]$, then $p$ is amplified by $R^{\epsilon
	\alpha }$ to offset the path loss, which is $R^{-\alpha }$, where $\alpha $
is the path loss exponent, with values greater than $2$. Reducing the power
of a UE as it gets closer to its serving BS is useful for two reasons. First
it saves the UE's battery. Second, it makes the UE less of an interferer to
UEs of neighbor cells. Combining the effects of FPC, power loss and fading,
the amount of power reaching the serving BS from the UE will be $%
pGR^{-\alpha (1-\epsilon )}$.

Referring to Figure \ref{fig:UL}, the RV $R_{\mathfrak{z}}$ is upper bounded by $U_{\mathfrak{z}}$, otherwise
the sample UE at $\mathfrak{z}$ would associate with the typical BS. Moreover, as can be seen, the
typical circle does not define an exclusion zone, as UEs can be located inside of it without breaking the associatiation rule. Let $I_{u}$ denote the interference caused by all the UEs, except the tagged, at the typical BS. Accordingly,
\begin{equation}
	I_{u}=\sum_{\mathfrak{z}\in \Psi \backslash \{ \mathfrak{u}\}}pG_{\mathfrak{z}%
	}R_{\mathfrak{z}}^{\alpha \epsilon }U_{\mathfrak{z}}^{-\alpha }
\label{Iu}
\end{equation}%
In the UL, the SIR of the typical BS, at distance $R$
from the tagged UE, is%
\begin{equation}
	\mathtt{SIR}_{\text{BS}}=\frac{pGR^{-\alpha (1-\epsilon )}}{I_{u}}
	\label{SIRu}
\end{equation}%
Consequently, the probability $p_{u}$ of UL coverage is 
\begin{equation*}
	p_{u}=\mathbb{P}[\mathtt{SIR}_{\text{BS}}>\xi ]\text{.}
\end{equation*}%
This probability can be visualized as being the average area or the average
fraction of users in coverage. As noted earlier, we perform analysis on a
randomly chosen BS assumed to be\ located at the origin associated with the
closest UE. Under assumption 1, the distribution of the distance $R$\ of the
closest mobile from the randomly chosen BS can be assumed Rayleigh given by (%
\ref{Reighlay}). This chosen BS is placed at the origin, and is henceforth
called the typical BS.

Referring to Figure \ref{fig:UL}, both $R$ and $R_{\mathfrak{z}}$  are Rayleigh
distributed, i.e. $f_{R}(r)=f_{R_{\mathfrak{z}}}(r)=2\lambda \pi
re^{-\lambda \pi r^{2}}$. Thus, $R_{\mathfrak{z}}$ ranges from $0$ to $R$,
with $R$ ranging from an arbitrarily small positive real number to $\infty $. Now, the conditional UL coverage probability is defined as%
\begin{eqnarray*}
	p_{u|R} &=&\mathbb{P}[\mathtt{SIR}_{\text{BS}}>\xi ] \\
	&=&\mathbb{E}\left[ \mathbb{P}\left[ \frac{pGR^{-\alpha (1-\epsilon )}}{I_{u}%
	}>\xi \right] \right] \\
	&=&\mathbb{E}\left[ \mathbb{P}\left[ G>\xi p^{-1}R^{\alpha (1-\epsilon
		)}I_{u}\right] \right] \\
	&\overset{(\text{a})}{=}&\mathbb{E}\left[ e^{-\xi p^{-1}I_{u}R^{\alpha
			(1-\epsilon )}}\right] \\
	&=&\mathcal{L}_{I_{u}}(\xi p^{-1}R^{\alpha (1-\epsilon )})
\end{eqnarray*}%
where $\mathcal{L}_{I_{u}}$ is the Laplace transform of the distribution of
the $I_{u}$ RV. In (a), we used the fact that $G\sim \exp (1)$, i.e. $%
f_{G}(x)=e^{-x}$, which implies that $\mathbb{P}[G>x]=e^{-x}$. Now, we
decondition on $R$, getting%
\begin{eqnarray}
	p_{u} &=&\int_{0}^{\infty }\left. \mathcal{L}_{I_{u}}(\xi p^{-1}R^{\alpha
		(1-\epsilon )})\right \vert _{R=r}f_{R}(r)dr  \notag \\
	&=&\int_{0}^{\infty }2\widetilde{\lambda }re^{-\widetilde{\lambda }r^{2}}%
	\mathcal{L}_{I_{u}}(\xi p^{-1}r^{\alpha (1-\epsilon )})dr  \label{pcMAIN}
\end{eqnarray}%
where $\widetilde{\lambda }=\pi \lambda $. We integrate from a point just
outside the origin, to skip the typical BS that resides there, to $\infty $
where the closest UE can possibly exist.

Next, we will embark on finding $\mathcal{L}_{I_{u}}$, the Laplace transform
of the distribution of the RV $I_{u}$.$\mathcal{\ }$Substituting for $I_{u}$
from (\ref{Iu}), gives 
\begin{eqnarray}
	\mathcal{L}_{I_{u}}(s) &=&\mathbb{E}[e^{-sI_{u}}]  \notag \\
	&=&\mathbb{E}\left[ \exp \left( s\sum_{\mathfrak{z}\in \Psi }-pG_{\mathfrak{z%
	}}R_{\mathfrak{z}}^{\alpha \epsilon }U_{\mathfrak{z}}^{-\alpha }\right) %
	\right]  \notag \\
	&=&\mathbb{E}\left[ \prod \limits_{\mathfrak{z}\in \Psi }\exp \left( -spG_{%
		\mathfrak{z}}R_{\mathfrak{z}}^{\alpha \epsilon }U_{\mathfrak{z}}^{-\alpha
	}\right) \right]  \label{EIz}
\end{eqnarray}%
In (\ref{EIz}), for each point $\mathfrak{z}\in \Psi $ there are three RVs: $%
G_{\mathfrak{z}},R_{\mathfrak{z}},U_{\mathfrak{z}}$. The $G_{\mathfrak{z}}$
are independent of the $R_{\mathfrak{z}}$ and of the $U_{\mathfrak{z}}$.
However, $U_{\mathfrak{z}}$ and $R_{\mathfrak{z}}$ are dependant in that $R_{%
	\mathfrak{z}}<U_{\mathfrak{z}}$ (Recall that $R_{z}$ is the distance between
an interfering UE $\mathfrak{z}$ and its typical BS, and $U_{\mathfrak{z}}$
is the distance between the same interfering UE $\mathfrak{z}$ and the
typical BS at the origin). That is $\mathbb{P}[R_{\mathfrak{z}}<x|U_{%
	\mathfrak{z}}=x]=1$, since if $U_{\mathfrak{z}}<R_{\mathfrak{z}}$ the
interfering UE $\mathfrak{z}$ would associate with the typical BS at the
origin.

We will next apply the expectation operator $\mathbb{E}$ to the three RVs
in (\ref{EIz}), one at a time, starting with the one that is independent of the
other two, namely $G_{\mathfrak{z}}$. Only \emph{one} of these expectations,
the one with respect to the PPP $\Psi $, will be evaluated using a PGFL and
the others using the standard definition of an expectation. Note that $\mathbb{E}_{G_{\mathfrak{z}}}$, $%
G_{\mathfrak{z}}\sim {Exp}(1)$, i.e. $f_{G_{\mathfrak{z}}}(x)=e^{-x}$. 
\begin{eqnarray}
	\mathcal{L}_{I_{u}}(s) &=&\mathbb{E}_{\Psi ,R_{\mathfrak{z}},G_{\mathfrak{z}%
	}}\left[ \prod \limits_{\mathfrak{z}\in \Psi }e^{-spG_{\mathfrak{z}}R_{%
			\mathfrak{z}}^{\alpha \epsilon }U_{\mathfrak{z}}^{-\alpha }}\right]   \notag
	\\
	\overset{(\text{a})}{=} &&\mathbb{E}_{\Psi ,R_{\mathfrak{z}}}\left[
	\prod \limits_{\mathfrak{z}\in \Psi }\mathbb{E}_{G_{\mathfrak{z}}}\left[
	e^{-spG_{\mathfrak{z}}R_{\mathfrak{z}}^{\alpha \epsilon }U_{\mathfrak{z}%
		}^{-\alpha }}\right] \right]   \notag \\
	\overset{(\text{b})}{=} &&\mathbb{E}_{\Psi ,R_{\mathfrak{z}}}\left[
	\prod \limits_{\mathfrak{z}\in \Psi }\int_{0}^{\infty }e^{-\left( 1+spR_{%
			\mathfrak{z}}^{\alpha \epsilon }U_{\mathfrak{z}}^{-\alpha }\right) x}dx%
	\right]   \notag \\
	&=&\mathbb{E}_{\Psi ,R_{\mathfrak{z}}}\left[ \prod \limits_{\mathfrak{z}\in
		\Psi }\frac{1}{1+spR_{\mathfrak{z}}^{\alpha \epsilon }U_{\mathfrak{z}%
		}^{-\alpha }}\right]   \label{LIus2}
\end{eqnarray}%
In (a) we used the fact that the $G_{\mathfrak{z}}$ are iid and in (b) we
used the fact that $f_{G_{\mathfrak{z}}}(x)=e^{-x}$.

Now, we consider the expectation with respect to $\Psi $, to uncondition on $%
U_{\mathfrak{z}}$, the distance between every point $\mathfrak{z}\in \Psi $
and the origin. We will use for this expectation a PGFL, since $U_{\mathfrak{%
		z}}$\ is distributed differently for each point $\mathfrak{z}\in \Psi $.
Based on the approximation we are proposing, the points of the
non-Poisson $\Psi $ PP are now relocated to the points of the (homogeneous)
Poisson $\Phi $ PP. In particular, each point $\mathfrak{z}\in \Psi $ will
be relocated to the position of the associated point $z\in \Phi $. That is,
we will consider each interfering UE at point $z\in \Phi $ emitting power $%
pR_{\mathfrak{z}}^{\alpha \epsilon }$, but causing interference with this same power at the typical BS, at a distance $D_{z}$ based on the UE relocation. This allows us to write%
\begin{equation*}
	\mathbb{E}_{\Psi }\left[ \prod \limits_{\mathfrak{z}\in \Psi }f(\mathfrak{z})%
	\right] \approx \mathbb{E}_{\Phi }\left[ \prod \limits_{z\in \Phi }f(z)\right]
	=e^{-\lambda \int_{\mathbb{R}^{2}}(1-f(x))}\text{.}
\end{equation*}%
Substituting for $f(y)$ from (\ref{LIus2}), converting to polar coordinates,
and substituting for the angle integral by $2\pi $, then%
\begin{eqnarray}
	\mathcal{L}_{I_{u}}(s) &=&\mathbb{E}_{R_{\mathfrak{z}}}\left[ \mathbb{E}%
	_{\Phi }\left[ \prod \limits_{z\in \Phi }\frac{1}{1+spR_{\mathfrak{z}%
		}^{\alpha \epsilon }D_{z}^{-\alpha }}\right] \right]  \notag\\
	&=&\mathbb{E}_{R_{\mathfrak{z}}}\left[ e^{-2\pi \lambda \int_{0}^{\infty }%
		\frac{1}{1+\left( sp\right) ^{-1}R_{\mathfrak{z}}^{-\alpha \epsilon
			}x^{\alpha }}xdx}\right] 
	\label{LIZ(s)}
\end{eqnarray}%
The integral includes the entire Euclidean space, so $r$ goes from $0$,
"very near" the origin to avoid having a BS at the origin where interference
is assessed, to infinity. However, we approximate this nearness by $0$, not
fearing a singularity in the denominator of the integrand since there is a $1$.

Note that $R_{\mathfrak{z}}$ is lower bounded by $U_{\mathfrak{z}}$, for if $R_{\mathfrak{z}}<U_{\mathfrak{z}}$, the UE at $\mathfrak{z}$ would associate with the typical BS. But note that $U_{\mathfrak{z}}$ has been
replaced now, through PP relocation, by $D_{\mathfrak{z}}$, which after
instantiation at the deconditioning step above becomes $x$. Therefore, when
we decondition on $R_{\mathfrak{z}}$ below, the integral will be from $0$ to $x$.

We will apply the last expectation, $\mathbb{E}_{R_{\mathfrak{z}}}$, using the
property that the expectation of an integral equals the integral
of the expectation of the integrand. The integral will extend from $R=r$ as
a minimum to $\infty $. Note that the distribution
of $R_{\mathfrak{z}}$ is Rayleigh, as it is an association distance (based on the closest distance).
In light of (\ref{LIZ(s)}), using the Rayleigh
distribution $f_{R_{\mathfrak{z}}}(y)=2\lambda \pi ye^{-\lambda \pi y^{2}}$,
we have 
\begin{eqnarray}
	\mathcal{L}_{I_{u}}(s) &=&e^{-2\pi \lambda \int_{0}^{\infty }\left( \mathbb{E%
		}_{R_{\mathfrak{z}}}\left[ \frac{1}{1+\left( sp\right) ^{-1}R_{\mathfrak{z}%
			}^{-\alpha \epsilon }x^{\alpha }}\right] \right) xdx}  \notag \\
	&=&e^{-2\pi ^{2}\lambda ^{2}\int_{0}^{\infty }x\int_{0}^{x^{2}}\frac{%
			e^{-\lambda \pi u}}{1+\left( sp\right) ^{-1}u^{-\alpha \epsilon /2}x^{\alpha
		}}dudx}  \label{LIus}
\end{eqnarray}%
where $u=y^{2}$. Recall that $D_{z}^{-\alpha }$ and $R_{\mathfrak{z}}$ are
dependent in that if $D_{z}=x$ then $R_{\mathfrak{z}}<x$. That is, the
distance $R_{\mathfrak{z}}$ between an interferer and the typical BS is
upper bounded by the distance $D_{z}$ between the (relocated) interfering UE
and the typical BS. From (\ref{Iu}) and (\ref{pcMAIN}), it follows that
\begin{eqnarray}
	\mathcal{L}_{I_{u}}(\xi p^{-1}r^{\alpha (1-\epsilon )}) &=&e^{-2\pi
		^{2}\lambda ^{2}\int_{0}^{\infty }x\int_{0}^{x^{2}}\frac{e^{-\lambda \pi u}}{%
			1+\left( sp\right) ^{-1}u^{-\alpha \epsilon /2}x^{\alpha }}dudx}  \notag \\
	&=&e^{-2\pi ^{2}\lambda ^{2}\int_{0}^{\infty }x\int_{0}^{x^{2}}\frac{\xi
			r^{\alpha (1-\epsilon )}e^{-\lambda \pi u}}{\xi r^{\alpha (1-\epsilon
				)}+u^{-\alpha \epsilon /2}x^{\alpha }}dudx}  \label{Liu}
\end{eqnarray}%
From (\ref{pcMAIN}) and (\ref{Liu}), we get 
\begin{equation}
	p_{u}=2\widetilde{\lambda }\int_{0}^{\infty }re^{-\widetilde{\lambda }%
		r^{2}}e^{-2\widetilde{\lambda }^{2}\xi r^{2\kappa (1-\epsilon
			)}\int_{0}^{\infty }x\int_{0}^{x^{2}}\frac{e^{-\widetilde{\lambda }u}}{\xi
			r^{2\kappa (1-\epsilon )}+u^{-\epsilon \kappa }x^{2\kappa }}dudx}dr
	\label{puWithLampda}
\end{equation}%
where $\widetilde{\lambda }=\pi \lambda $ and $\kappa =\alpha /2$.

From (\ref{puWithLampda}), it \emph{appears} that the UL coverage probability $p_{u}
$ is dependent on the BS density $\lambda $. Indeed, this belief is popular
research work on stochastic geometric models of the cellular UL system (See,
for example, \cite{ElSawy17, Chun20, Haroon20, Herath18, Kouzayha20, Mariam21, Andrews16, Jia19, Gao19, Ali19, Kundu20, Sadeghabadi20, Wang19}). However, we
will show in Theorem 2 below that this belief is flawed. Specifically, we will show that the presence of $\lambda $ in (\ref{puWithLampda}), as in a large number of similar
expressions in the literature of stochastic geometric modelling of cellular
networks, is superfluous. 

\begin{description}
	\item \textbf{Theorem 2}: Under the stochastic geometric model of the cellular UL
	system, the UL coverage probability $p_{u}$ is independent of the BS density 
	$\lambda $.
\end{description}

\textbf{Proof}: The proof is attained through a sequence of changes of
variables. Starting with (\ref{puWithLampda}), use the substitution $v=r^{2}$
to get%
\begin{equation*}
	p_{u}=\widetilde{\lambda }\int_{0}^{\infty }e^{-\widetilde{\lambda }v}e^{-2%
		\widetilde{\lambda }^{2}\xi v^{\kappa (1-\epsilon )}\int_{0}^{\infty
		}x\int_{0}^{x^{2}}\frac{e^{-\widetilde{\lambda }u}}{\xi v^{\kappa
				(1-\epsilon )}+u^{-\epsilon \kappa }x^{2\kappa }}dudx}dv
\end{equation*}%
Use $y=x^{2}$ to get%
\begin{equation*}
	p_{u}=\widetilde{\lambda }\int_{0}^{\infty }e^{-\widetilde{\lambda }v}e^{-%
		\widetilde{\lambda }^{2}\xi v^{\kappa (1-\epsilon )}\int_{0}^{\infty
		}\int_{0}^{y}\frac{e^{-\widetilde{\lambda }u}}{\xi v^{\kappa (1-\epsilon
				)}+u^{-\epsilon \kappa }y^{\kappa }}dudy}dv
\end{equation*}%
Use $x=\widetilde{\lambda }u$ to get%
\begin{equation*}
	p_{u}=\widetilde{\lambda }\int_{0}^{\infty }e^{-\widetilde{\lambda }v}e^{-%
		\widetilde{\lambda }\xi v^{\kappa (1-\epsilon )}\int_{0}^{\infty }\int_{0}^{%
			\widetilde{\lambda }y}\frac{e^{-x}}{\xi v^{\kappa (1-\epsilon )}+\left( 
			\frac{x}{\widetilde{\lambda }}\right) ^{-\epsilon \kappa }y^{\kappa }}dxdy}dv
\end{equation*}%
Use $z=\widetilde{\lambda }v$ to get

\begin{equation*}
	p_{u}=\int_{0}^{\infty }e^{-z}e^{-\widetilde{\lambda }\xi z^{\kappa
			(1-\epsilon )}\int_{0}^{\infty }\int_{0}^{\widetilde{\lambda }y}\frac{e^{-x}%
		}{\xi z^{\kappa (1-\epsilon )}+x^{-\epsilon \kappa }\left( \widetilde{%
				\lambda }y\right) ^{\kappa }}dxdy}dz
\end{equation*}%
Finally, use $u=\widetilde{\lambda }y$ to get

\begin{equation}
	p_{u}=\int_{0}^{\infty }e^{-z\left( 1+\xi z^{\kappa (1-\epsilon
			)-1}\int_{0}^{\infty }\int_{0}^{u}\frac{e^{-x}}{\xi z^{\kappa (1-\epsilon
				)}+x^{-\epsilon \kappa }u^{\kappa }}dxdu\right) }dz  
	\label{puWithoutLampda}
\end{equation}%
where $\lambda $ has totally disappeared, proving the theorem. $\blacksquare$

This is the final expression for the UL coverage probability. The striking
observation about the above $p_{u}$ is that for all $\alpha $, the UL
coverage probability is independent of the density $\lambda $ of the BSs. A possible interpretation is that as the BS density gets higher, the Voronoi cells get smaller, and the UEs get closer to their serving BSs, and vice versa. This means that the increase in interference due to the increase of BS density, will be offset by an increase in signal due to the nearness of the tagged UE to the typical BS, keeping SIR, hence the coverage probability, unchanged.

Just like Theorem 1, Theorem 2 calls for a revisit to the UL coverage probability expressions (and their derivatives)
that include $\lambda $. Those expressions exist in a huge number of articles, such as those cited just above Theorem 2. It is worth examining them for possible elimination of a parameter that may exist misleadingly. 

\textbf{UL Special cases}:

First, we will consider the special case of $\epsilon =0$, i.e. there is no FPC by the UE. For this case, we get from (\ref{puWithoutLampda}) that

\begin{eqnarray}
	p_{u} &=&\int_{0}^{\infty }e^{-z\left( 1+\xi z^{\kappa -1}\int_{0}^{\infty
		}\int_{0}^{u}\frac{e^{-x}}{\xi z^{\kappa }+u^{\kappa }}dxdu\right) }dz 
	\notag \\
	&=&\int_{0}^{\infty }e^{-z\left( 1+\xi z^{\kappa -1}\int_{0}^{\infty }\frac{%
			1-e^{-u}}{\xi z^{\kappa }+u^{\kappa }}du\right) }dz  \label{puk}
\end{eqnarray}
Using a computer algebra system (e.g., Maxima (R)) or a table of integrals
(e.g., \cite{Gradshteyn15}: p. 325, \#3.241], one can find%
\begin{equation*}
	\int_{0}^{\infty }\frac{1}{\xi z^{\kappa }+u^{\kappa }}du=\frac{1}{\kappa
		\xi ^{1-1/\kappa }z^{\kappa -1}}B\left( 1-\frac{1}{\kappa },\frac{1}{\kappa }%
	\right) ,\qquad 0<\frac{1}{\kappa }<1
\end{equation*}%
where $B(x,y)$ is the Bessel function of $x$ and $y$. Substituting in (\ref%
{puk}), we get%
\begin{equation*}
	p_{u}=\int_{0}^{\infty }e^{-z\left( 1+\left( \frac{\sqrt[\kappa ]{\xi }}{%
			\kappa }B\left( 1-\frac{1}{\kappa },\frac{1}{\kappa }\right) -\xi z^{\kappa
			-1}\int_{0}^{\infty }\frac{e^{-u}}{\xi z^{\kappa }+u^{\kappa }}du\right)
		\right) }dz
\end{equation*}
For $\alpha =4$ ($\kappa =2$), one can find
\begin{equation*}
	\int_{0}^{\infty }\frac{1}{\xi y^{2}+x^{2}}dx=\frac{1}{2y\sqrt{\xi }}B\left(
	0.5,0.5\right) =\frac{\pi }{2y\sqrt{\xi }}
\end{equation*}%
Substituting in (\ref{puk}), we get%
\begin{equation}
	p_{u}=\int_{0}^{\infty }e^{-z\left( 1+\left( \frac{\sqrt{\xi }}{2}B\left( 
		\frac{1}{2},\frac{1}{2}\right) -\xi z\int_{0}^{\infty }\frac{e^{-u}}{\xi
			z^{2}+u^{2}}du\right) \right) }dz  \label{puk=2}
\end{equation}%
Using a computer algebra system (e.g., Maxima (R)) or a table of integrals
(e.g., \cite{Gradshteyn15}: p. 343, \#3.354], one can find the integral at
the exponent%
\begin{equation}
	\int_{0}^{\infty }\frac{e^{-u}}{\xi z^{2}+u^{2}}dx=\frac{1}{\sqrt{\xi }z}%
	\left[ \mathop{\rm ci}(\sqrt{\xi }z)\sin (\sqrt{\xi }z)-\mathop{\rm si}(%
	\sqrt{\xi }z)\cos (\sqrt{\xi }z)\right]   \label{si}
\end{equation}%
where%
\begin{equation*}
	\mathop{\rm si}(x)=\mathop{\rm Si}(x)-\frac{\pi }{2}=-\int_{x}^{\infty }%
	\frac{\sin t}{t}dt
\end{equation*}%
and%
\begin{equation*}
	\mathop{\rm ci}(x)=\mathop{\rm Ci}(x)=-\int_{x}^{\infty }\frac{\cos t}{t}dt
\end{equation*}%
are the sine and cosine integrals, respectively. Using (\ref{si}) and the
fact that $B\left( \frac{1}{2},\frac{1}{2}\right) =\pi $ in (\ref{puk=2}),
the coverage probability for $\alpha =4$ becomes%
\begin{equation}
	p_{u}=\int_{0}^{\infty }e^{-z\left( 1+\sqrt{\xi }\left( \frac{\pi }{2}-%
		\mathop{\rm ci}(\sqrt{\xi }z)\sin (\sqrt{\xi }z)+\mathop{\rm si}(\sqrt{\xi }%
		z)\cos (\sqrt{\xi }z)\right) \right) }dz  \label{puAlpha=4}
\end{equation}

Second, we will consider the special case of $\epsilon =1$, i.e. there is channel inversion by the UE. For this case, we get from (\ref{puWithoutLampda}) that
\begin{eqnarray*}
	p_{u} &=&\int_{0}^{\infty }e^{-z\left( 1+\xi z^{-1}\int_{0}^{\infty
		}\int_{0}^{u}\frac{x^{\kappa }e^{-x}}{\xi x^{\kappa }+u^{\kappa }}%
		dxdu\right) }dz \\
	&=&\int_{0}^{\infty }e^{-z}e^{-\xi \int_{0}^{\infty }\int_{0}^{u}\frac{%
			x^{\kappa }e^{-x}}{\xi x^{\kappa }+u^{\kappa }}dxdu}dz \\
	&=&\exp \left( -\xi \int_{0}^{\infty }\int_{0}^{u}\frac{x^{\kappa }e^{-x}}{%
		\xi x^{\kappa }+u^{\kappa }}dxdu\right)
\end{eqnarray*}

At $\kappa =2$: 
\begin{equation*}
	p_{u}=\exp \left( -\xi \int_{0}^{\infty }\int_{0}^{u}\frac{x^{2}e^{-x}}{\xi
		x^{2}+u^{2}}dxdu\right) 
\end{equation*}%
which could not be simplified further because the inner integral could not
be evaluated.

\section{Simulation}
\label{SIM}
Despite diligent search in the literature, we have not spotted
any formal stochastic geometric simulation algorithms. What is 
available is only code, and even that is rare, poorly documented, and sporadic. 
In this section, we provide the simulation algorithms we have developed for the present
article. Although the algorithms were primarily intended to validate our
analytical results, we feel they can be of value in
similar stochastic geometric research. Since DL and UL are different, as we
have seen in the modelling, their simulation algorithms are different, and
therefore will be introduced separately. However, we will give below some
points common to both.

Although we have proved that the BS density $\lambda $ has no bearing on the
coverage probability in either DL or UL, we have included it in the two
simulation algorithms. The reason for that is to allow the interested reader
to verify practically what we have proven theoretically. Indeed, one can use
the simulation algorithms below to ascertain that for any value of $\lambda $%
, the coverage probability remains the same for the same path loss exponent $%
\alpha $ and SIR threshold $\xi $. Actually, it was experimentation of this
sort that led us to discover the irrelevance of $\lambda $ and motivated the
theoretical proof presented earlier.

\subsection{Downlink}

The modelling of the DL mode was undoubtedly simple. Its simulation is just
as simple. After scattering the BSs of a realization in the simulation
window, we select the closest BS to the origin and make it the tagged BS,
considering that the typical UE is at the origin already. That is, we do not
need to generate any UEs. The transmission received at the origin from the
tagged BS will be then considered signal and every other transmission
received there considered interference. It is a simple matter then to
calculate the SIR at the typical UE and use it to determine if the typical
UE is covered by the tagged BS for this realization, for the given $\xi $.
The coverage probability is evaluated after all $\mathcal{N}$ realizations
have been generated, by dividing the number of realizations where the
typical UE was covered by the total number $\mathcal{N}$ of realizations.
The algorithm then proceeds to calculate the SIR at the typical UE, using (%
\ref{SIRd}).

\subsection{Uplink}

UL simulation is more involved than DL simulation, as was the case with
their modelling. A key element for a successful UL simulation is the
association table, $A$, an example of which is shown in Table \ref{table:A}.
For a network of $N$ BSs, this $N\times 2$ table has one column for the BSs
and another for the UEs. In particular, the table stores the coordinates of
each BS and those of its associated UE (or conversely, each UE and its
serving BS) in one row. As such, the table not only tells which UE is
associated to which BS, but also allows calculating the distance between any
BS or UE and any other BS or UE. These distances are crucial to calculate
the interference at a given receiver. For example, referring to Table \ref%
{table:A}, the first BS is at the origin and its associated UE is $734.2$ m
west and $628.4$ m north, i.e. the BS and UE are $\sqrt{734.2^{2}+628.4^{2}}%
=966.4$ m apart. We note that the third BS has no associated UE yet.


\begin{table}[H]
	\caption{Example BS-UE Association table $A$ for UL simulation.} 
	\centering 
	\begin{tabular}{c c } 
		\hline\hline 
		BS & UE\\ 
		\hline 
		(0,0) & (-734.2,628.4)\\
		(1243.2,-221.4) & (973.2,1628.4)\\
		(-345.2,928.4) & NULL \\
		$\ldots$ & $\ldots$\\
		\hline 
	\end{tabular}
	\label{table:A} 
\end{table}

The association table $A$ is built by the simulation algorithm as follows.
First, the BS column is filled at once by generating $N$ pairs of random
numbers (drawn from a uniform distribution) in the interval $[-S,S]$, where $%
S$ is the simulation window (square) side. The first number in the pair is
taken as the $x$ coordinate, and the second as the $y$ coordinate. Then the
UEs are added incrementally, from top to bottom, one at a time based on the
BS-UE association rule, which says:\ a UE must be closer to its serving BS
than to any other BS in the cellular network. Adding the first UE to the
table is simplest. A pair of random numbers is first generated (drawn from a
uniform distribution) in the interval $[-S,S]$ to determine the location of
the UE in the simulation window. Then the distances between this UE and all
the BSs in the network, whose locations are in column $1$ of the table, are
calculated. The UE is assigned immediately to the closest BS.

\begin{algorithm}
	
	\SetKwInOut{Input}
	\SetKwInOut{\textbf{Input}: $\mathcal{N}$, $\xi$, $p$, $\lambda$, $S$, $\alpha$}\\
	\SetKwInOut{Output}
	\SetKwInOut{\textbf{Output}: $p_d$}\\
	Covered  := 0\\
	\For{$i:= 1$ \textnormal{to} $\mathcal{N}$}
	{
		\tcc{Find number $N$ of BSs to deploy in simulation window:}
		Generate a Possion distributed random number $N$, $N \sim$ Pois($\lambda S^2$) \\
		
		\tcc{Scatter $N$ BSs over an $S \times S$ square window, storing the location at a 1 column array $D$:}
		
		\For{$i: = 1$ \textnormal{to} $N$} 	{
			
			Generate two uniformly distributed random numbers $(x,y)$, where $x,y \sim$ U($-S,S$), for BS $i$ location.\\
			$D(i)=(x,y)$
		}
		
		Identify BS $n$ nearest the origin and designate it tagged BS\\		
		
		\tcc{Find SIR at typical UE and check if the latter is covered by tagged BS for the given $\xi$:}		
		
		Generate a Poisson random number $G$, where $G \sim$ Pois(1) \\	
		Calculate distance $R$ between the tagged BS and origin\\
		NUM $= p G R^{-\alpha}$ \\	
		DENUM $= 0$ \\	
		\For {$i:=1$ \textnormal{to} $N$}
		{
			\If{$i \ne n$}
			{ 
				Generate a Poisson random number $G_i$, where $G_i \sim$ Pois(1) \\	
				Use array $D$ to calculate the distance $D_i$ between BS $i$ and origin\\
				DENUM $=$ DENUM $+ p G_i {D_i}^{-\alpha}$\\
			}	
			
		}
		SIR := NUM/DENUM \\
		
		\If{\textnormal{SIR} $> \xi$}
		{ 
			Covered  :=   Covered + 1
		}
	} 
	\tcc{Calculate DL coverage probability $p_d$ for this threshold $\xi$:}
	$p_d$  := Covered $/ \mathcal{N}$
	
	\caption{Simulation of cellular DL system}
\end{algorithm}
\label{alg:DL}
\hspace{2cm}

\begin{algorithm}
	
	\SetKwInOut{Input}
	\SetKwInOut{\textbf{Input}: $\mathcal{N}$, $\xi$, $p$, $\lambda$, $S$, $\alpha$, $\epsilon$}\\
	\SetKwInOut{Output}
	\SetKwInOut{\textbf{Output}: $p_u$}\\
	Covered  := 0\\
	\For{$i:= 1$ \textnormal{to} $\mathcal{N}$}
	{
		\tcc{Find number $N$ of BSs to deploy in simulation window:}
		Generate a Possion distributed random number $N$, $N \sim$ Pois($\lambda S^2$) \\
		
		\tcc{Build table $A$ for a  network of $N$ BSs and $N$ UEs:}
		
		\For{$i: = 1$ \textnormal{to} $N$} 	{
			$A(i,1)=A(i,2)=$ NULL 
			\tcc{Empty the association table, $A$}
		}
		
		\tcc{1- Fill in column 1 of $A$ with BSs, making BS 1 typical.}
		
		$A(1,1)=(0,0)$							
		
		\For{$i: = 2$ \textnormal{to} $N$} 	{
			
			Generate random location $(x,y)$, where $x,y \sim$ U($-S,S$), for BS $i$.\\
			$A(i,1)=(x,y)$
		}
		
		\tcc{2- Fill in column 2 of table $A$ with UEs.}								
		
		\While{\textnormal{($\exists$ \textnormal{NULL} $\in A$)}}
		{
			Generate random location $(u,v)$, where $u,v \sim$ U($-S,S$), for a UE.\\
			
			Calculate the $N$ distances between the UE and every BS $i$\\
			Identify BS $n$, the BS nearest the UE\\
			\If{\textnormal{$A(n,2)=$\textnormal{NULL}}}
			{ 
				$A(n,2)=(u,v)$
				\tcc{Assocate UE with BS $n$ if available.}
			}
		}    
		
		\tcc{Find SIR at typical BS, update `Covered' \& calculate $p_u$:}		
		
		Generate a Poisson random number $G$, where $G \sim$ Pois(1) \\	
		Calculate from table $A$ distance $R$ between typical BS and tagged UE\\
		NUM $= p G R^{-\alpha(1-\epsilon)}$ \\	
		DENUM $= 0$ \\	
		\For {$i:=2$ \textnormal{to} $N$}{
			Generate a Poisson random number $G_i$, where $G_i \sim$ Pois(1) \\	
			Calculate from table $A$ distance $R_i$ between UE $i$ and serving BS\\
			Calc. from table $A$ distance $U_i$ between UE $i$ and typical BS\\
			DENUM $=$ DENUM $+ p G_i {R_i}^{\alpha \epsilon} {U_i}^{-\alpha}$\\
		}
		SIR := NUM/DENUM \\
		
		\If{\textnormal{SIR} $> \xi$}
		{ 
			Covered  :=   Covered + 1
		}
	} 
	$p_u$  := Covered $/ \mathcal{N}$
	
	\caption{Simulation of cellular UL system}
	\label{alg:UL}
\end{algorithm}

The addition of the second UE gets a little harder. First, we generate the
location and calculate the distance as we did for the first UE. However, now
there is a possibility that the second UE is closest to the BS to which the
first UE was associated. Here the attempt fails. We keep making attempts
till we find a nearest BS that is available, to which we make the
association by recording the coordinates of the UE in the same row of that
nearest BS. Of course, adding UEs keeps getting harder and harder as the
number of associations increases since the chance of finding a BS that is
both closest to the UE and at the same time available becomes increasingly
small. For this probabilistic behavior, building the association table is
the most time consuming part of the algorithm. All the other parts are just
fast computations based on distances obtained from the numbers of this table.

The algorithm then proceeds to calculate the SIR at the typical BS, using (%
\ref{SIRu}). Note that in the simulation we can calculate exactly the two
distances: $R_{\mathfrak{z}}$, the distance between the interfering UE and
its serving BS, and $U_{\mathfrak{z}}$, the distance between the interfering
UE and the typical BS. That is, we can do in an exact way in the simulation,
what we did in an approximate way in the modelling. Interestingly, however,
the results of both the modelling and simulation conform impressively,
testifying to the validity of the approximation made in the modelling.

Assessing the time complexity of both algorithms is difficult since the
number $N$ of BSs, i.e. PPP points, generated in each realization is not
constant, but a Poisson distributed RV ranging from $0$ to $\infty $. This
difficulty increases even more in the UL case, where the way the association
table $A$ is built is also random, as explained earlier.

\section{Experimental Work}
\label{EW}
We implemented the simulation algorithms in the Matlab (R) language as two
separate programs, one for DL and the other for UL. We coded the algorithms
in Matlab (R). The simulation time took around 1 hr on a Laptop of $8$ GB
RAM and Core $5$ CPU at $2.4$ GHz, for $3000$ runs, on a simulation window
in the form of a square of side $2000$ m. Matlab (R) was also used to
perform the computations of the analytical results, and proved particularly
efficient with numerical integration of formidable functions.

An important implementation tip is in order. The pseudo code shown in the two
simulation algorithms calculates the coverage probability for only one value
of the threshold, $\xi $. To sketch a smooth curve over a reasonable range
of the threshold, such as that of the coverage probability figures below, where the range is from $-15$ dB to $15$ dB, we should
then run the algorithm $\mathcal{N}$ times, with $\mathcal{N}$ typically
exceeding $3000$ for good convergence, for each of the $31$ dB values. To
save simulation time, however, we did something interesting that yielded the
same results nonetheless. In each realization, we tested the resulting SIR
with all the values of $\xi $ in the range, starting from $-15$ dB and going
upwards, incrementing the variable `Covered' each time the former exceeds
the latter for this particular value of $\xi $, for which an array, rather
than a single variable, is created. Once the former stops exceeding the
latter, we quit the comparisons for this realization. This trick reduces the
simulation time significantly, yet produces the same results. The same
thing, by the way, applies in UL when using more than one value of the power
control factor, $\epsilon $. Instead of running $\mathcal{N}$ realizations
for each $\epsilon $, we use the same realization to get results for as many 
$\epsilon $ values as desired. Of course, one would use arrays for these
multiple results, instead of the variables in the algorithms below which are
intended for only one $\xi $ value and one $\epsilon $ value (in case of
UL.) We always calculate the SIR at the typical receiver placed at the
origin (which is a UE in DL and a BS in UL.)

From our experience, there are two tricks that can make the running of the
algorithms extremely fast. First, we use the same realization to check for
all required thresholds. For example, in our experiments we use 31
thresholds, from $-15$ dB to $15$ dB, in increments of $1$ dB. Then, when we
obtain the SIR for a realization, we keep checking (using a loop) it against
the thresholds from the smaller going towards the bigger, within the same
realization. Once the SIR fails to exceed a threshold, we exit (or break)
the loop, since there is no point in checking the SIR against greater
thresholds. Moreover, in the UL system in particular, we use also the same
realization for all the values of the power control factor $\epsilon $ under
consideration. In some implementations we have seen, there is one
realization per threshold value per control factor value, totalling $93$
realizations if there are $31$ threshold values and $3$ power factor values.
These $93$ realizations are replaced by only $1$ realization, a huge savings
in terms of simulation time.

Furthermore, in our UL algorithm we used a trick in building the association
table that accelerated the simulation immensely. Unlike popular
implementations which insist on building the association table strictly from
top down, starting by associating the first BS, then the second BS, and so
on, we associate with the BS that turns out to be closest to the UE just
generated, provided the BS is available. We note in passing that we start in
UL by placing one BS at the origin, designating it to be the typical BS.
This BS can in fact be at any row of the association table $A$, so it might
as well be at the first row, for easy reference.


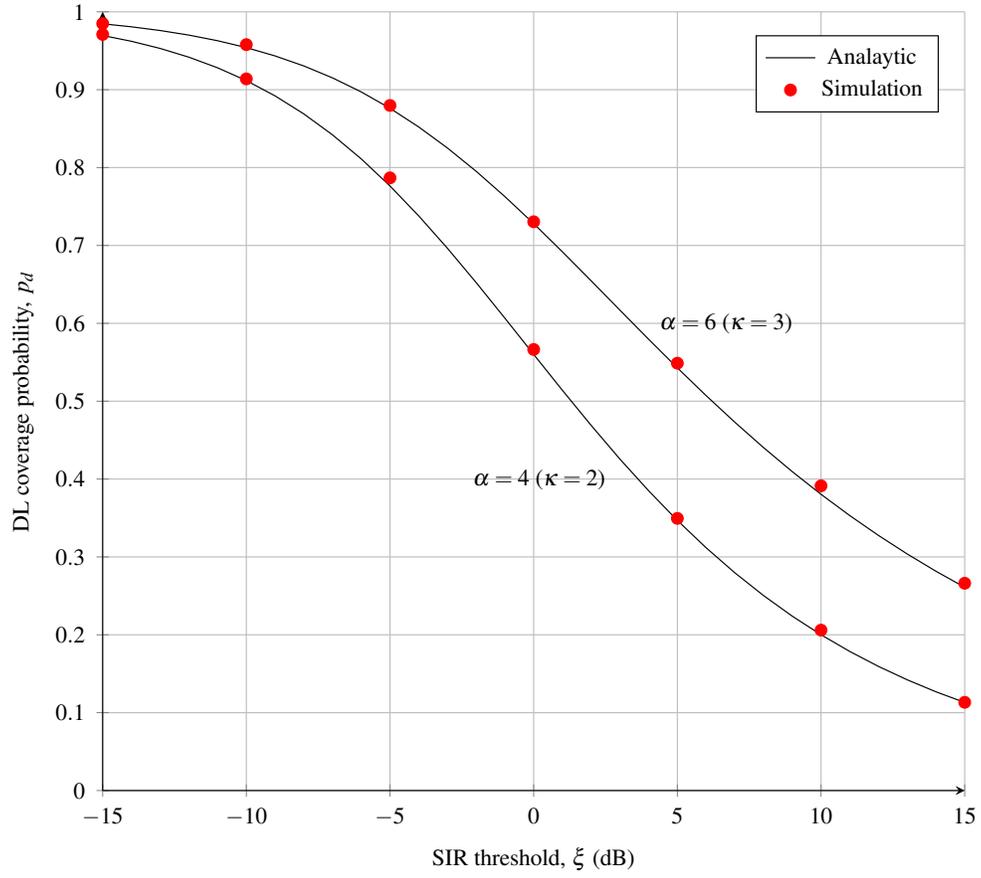
\begin{figure}[H]
	\begin{center}
		
		\begin{tikzpicture}[scale=1.1]
			\begin{axis}[
				xlabel={\scriptsize{SIR threshold, $\xi$ (dB)}} ,ylabel={\scriptsize{DL coverage probability, $p_d$ }},
				height=11cm,
				width=12cm,
				grid=major,
				axis lines=left,
				ymax=1,
				ymin=0,
				y tick label style={
					font=\scriptsize},
				xmin=-15,
				xmax=15,
				x tick label style={ 
					font=\scriptsize},
				xscale=1,
				legend pos=north east,
				]
				
				\addplot [no marks,mark= *,  color=black] coordinates {
					
					(-15,0.96965)
					(-14,0.96219)
					(-13,0.95301)
					(-12,0.94178)
					(-11,0.92814)
					(-10,0.9117)
					(-9,0.89208)
					(-8,0.86894)
					(-7,0.842)
					(-6,0.81113)
					(-5,0.77636)
					(-4,0.73793)
					(-3,0.69632)
					(-2,0.65223)
					(-1,0.6065 )
					(0,0.5601)
					(1,0.51396)
					(2,0.46895)
					(3,0.4257)
					(4, 0.38499)
					(5,0.34694)
					(6,0.3118)
					(7,0.27963)
					(8,0.25038)
					(9,0.22391)
					(10,0.20005)
					(11,0.17861)
					(12,0.15939)
					(13,0.14219)
					(14,0.12681)
					(15,0.11308)
					
				};
				\addlegendentry{\scriptsize{Analaytic}}
				\addplot [only marks,mark size=2pt,  color=red] coordinates {
					(-15, 0.971)
					(-10,0.9139)
					(-5,0.7868)
					(0,0.5664)
					(5,0.3495)
					(10,0.206)
					(15,0.1133)
				};
				\addlegendentry{\scriptsize{Simulation}}
				
				\addplot [no marks,mark= *,  color=black] coordinates {
					
					(-15,0.98462)
					(-14,0.98078)
					(-13,0.97602)
					(-12,0.97014)
					(-11,0.96291)
					(-10, 0.95409)
					(-9,0.94339)
					(-8,0.93052)
					(-7,0.91519)
					(-6,0.89714 )
					(-5,0.87617)
					(-4,0.85215)
					(-3,0.82511)
					(-2,0.79519)
					(-1,0.76268)
					(0,0.72804)
					(1,0.6918)
					(2,0.65457)
					(3,0.61697)
					(4,0.57955)
					(5, 0.54283)
					(6, 0.50721)
					(7,0.47299)
					(8,0.44037)
					(9,0.40949)
					(10,0.3804)
					(11,0.35311)
					(12,0.32759)
					(13,0.30378)
					(14,0.28161)
					(15,0.26099)
				};
				
				\addplot [only marks,mark size=2pt,  color=red] coordinates {
					(-15, 0.9849)
					(-10,0.9579)
					(-5,0.8798)
					(0,0.7304)
					(5,0.5488)
					(10,0.3913)
					(15,0.2662)
				};
				\node [right] at (axis cs:    4, 0.6) {{{\scriptsize{\color{black} $\alpha=6$ ($\kappa=3$)}}}};
				\node [right] at (axis cs:    -2.5, 0.4) {{\scriptsize{\color{black} $\alpha=4$ ($\kappa=2$)}}};
			\end{axis}
		\end{tikzpicture}
		\caption[state,]{DL coverage probability, $p_d$, for two values of $\alpha$}
		\label{fig:DLP}
	\end{center}
\end{figure}


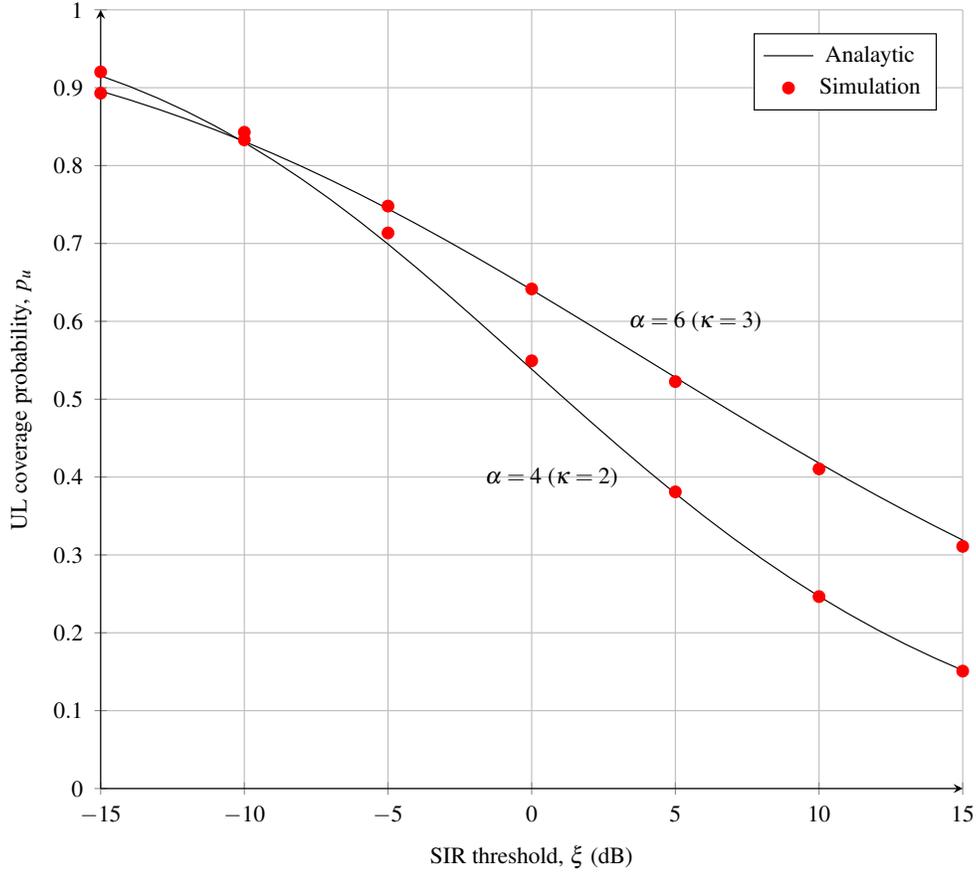
\begin{figure}[H]
	\centering
	\begin{center}
		
		\begin{tikzpicture}[scale=1.1]
			
			\begin{axis}[
				xlabel={\scriptsize{SIR threshold, $\xi$ (dB)}} ,ylabel={\scriptsize{UL coverage probability, $p_u$ }},
				height=11cm,
				width=12cm,
				grid=major,
				axis lines=left,
				ymax=1,
				ymin=0,
				y tick label style={
					font=\scriptsize},
				xmin=-15,
				xmax=15,
				x tick label style={ 
					font=\scriptsize},
				xscale=1,
				legend pos=north east,
				]
				
				\addplot [no marks,mark= *,  color=black] coordinates {
					
					(-15,0.89563)
					(-14,0.88447)
					(-13,0.87243)
					(-12,0.85951)
					(-11,0.84569)
					(-10,0.83097)
					(-9,0.81535)
					(-8,0.79886)
					(-7,0.78152)
					(-6,0.76337)
					(-5,0.74444)
					(-4,0.72481)
					(-3,0.70451)
					(-2,0.68363)
					(-1,0.66225)
					(0,0.64043)
					(1,0.61828)
					(2,0.59587)
					(3,0.57329)
					(4,0.55064)
					(5,0.52801)
					(6,0.50547)
					(7,0.48312)
					(8,0.46103)
					(9,0.43928)
					(10,0.41792)
					(11,0.39703)
					(12,0.37665)
					(13,0.35683)
					(14,0.33761)
					(15,0.31903)
					
				};
				\addlegendentry{\scriptsize{Analaytic}}
				\addplot [only marks,mark size=2pt,  color=red] coordinates {
					(-15, 0.893)
					(-10,0.833)
					(-5,0.748)
					(0,0.6416)
					(5,0.5226)
					(10,0.4105)
					(15,0.311)
				};
				\addlegendentry{\scriptsize{Simulation}}
				
				\addplot [no marks,mark= *,  color=black] coordinates {
					(-15,0.91528)
					(-14, 0.9017)
					(-13,0.88642)
					(-12,0.86937)
					(-11,0.85047)
					(-10, 0.8297)
					(-9,0.80705)
					(-8,0.78257)
					(-7,0.75634)
					(-6,0.72848)
					(-5,0.69915)
					(-4,0.66856)
					(-3,0.63694)
					(-2,0.60455)
					(-1,0.57165)
					(0,0.53853)
					(1,0.50547)
					(2,0.47273)
					(3,0.44057)
					(4,0.40922)
					(5, 0.37886)
					(6, 0.34968)
					(7,0.32179)
					(8,0.29531)
					(9,0.27031)
					(10,0.24682)
					(11,0.22486)
					(12,0.20442)
					(13,0.18547)
					(14,0.16797)
					(15,0.15187)
				};
				
				\addplot [only marks,mark size=2pt,  color=red] coordinates {
					(-15, 0.9203)
					(-10,0.8428)
					(-5,0.7135)
					(0,0.5493)
					(5,0.381)
					(10,0.2465)
					(15,0.1508)
				};
				\node [right] at (axis cs:    3, 0.6) {{\scriptsize{\color{black} $\alpha=6$ ($\kappa=3$)}}};
				\node [right] at (axis cs:    -2, 0.4) {{\scriptsize{\color{black} $\alpha=4$ ($\kappa=2$)}}};
			\end{axis}
			
		\end{tikzpicture}
		\caption[state]{UL coverage probability, $p_u$, for two values of $\alpha$, and with no FPC, i.e. $\epsilon=0$. }
		\label{fig:ULP}
		
	\end{center}
\end{figure}

We sketch in Figures \ref{fig:DLP} and \ref{fig:ULP} the analytical and
simulation values of the DL and UL coverage probabilities, respectively, for
a range of a threshold, $\xi $, extending from $-15$ dB to $15$ dB. For Figure \ref{fig:DLP}, we used (\ref{pd4})  and (\ref{pd6}) to sketch the analytical results for the cases of $\alpha=4$ and $\alpha=6$, respectively, and Aglorithm 1 to sketch the simulation results. On the other hand, for Figure \ref{fig:ULP}, we used (\ref{puWithoutLampda})  and (\ref{puAlpha=4}) to sketch the analytical results for the cases of $\alpha=6$ and $\alpha=4$, respectively, and Algorithm 2 to sketch the simulation results. The UL coverage probability is sketched for the case of no FPC, i.e. $\epsilon=0$.

As can
be seen the agreement between the two types of values is excellent. From the
actual values we used for graphig the curves, the difference between the
analytical and simulation results is within only $1\%$, out of as few as $%
3000$ simulation runs (realizations) on a $2000\times 2000$ m simulation
window.

\section{Conclusions}
\label{CONC}
In this article we have investigated thoroughly the modelling and simulation
of cellular DL and UL cellular channels using stochastic geometry. A number
of contributions have been made in the process. First, we introduced the DL
and UL models concisely together so as to expose their similarities and
differences vividly. In addition, we were able to derive closed form
expressions for the DL coverage probability in two special cases. Then, for
UL, known to be challenging, we introduced an approximation that
circumvented the challenge and yet yielded excellent results validated later
by simulation. For UL also, we obtained for the coverage probability an
expression, though not closed form, but simple enough to calculate easily.
For the simulation dimension of our study, we presented two effecient Monte
Carlo simulation algorithms designed to validate the models, but can be
useful to anyone using stochastic geometric in the communications field or
otherwise. Finally, we have proved two theorems that go against established
belief in the cellular modelling community. Namely, we have proved that under the
stochastic geometric model, the coverage probability in either DL and UL is
independent of the BS density. This finding calls for a revisit to a large
body of results published in the past decade with the BS density present superfluously in them.

\end{document}